\documentclass{jfm}
\usepackage{graphicx}
\usepackage{epstopdf, epsfig}

\usepackage{xcolor}
\usepackage{amsmath}

\shorttitle{On the initiation and sustenance of flow-induced vibrations}
\shortauthor{K. Menon and R. Mittal}

\title{On the initiation and sustenance of flow-induced vibration of cylinders: insights from force partitioning}

\author{Karthik Menon\aff{1}
  \corresp{\email{kmenon@jhu.edu}},
 \and Rajat Mittal\aff{1}
 \corresp{\email{mittal@jhu.edu}}}

\affiliation{\aff{1}Department of Mechanical Engineering, Johns Hopkins University, Baltimore, MD 21218, USA}

\begin{document}

\maketitle

\begin{abstract}
The focus of this work is to dissect the physical mechanisms that drive and sustain flow-induced, transverse vibrations of cylinders. The influence of different mechanisms is quantified by using a method to partition the fluid dynamic force on the cylinder into distinct, physically relevant components. In conjunction with this force partitioning, calculations of the energy extracted by the oscillating body from the flow are used to make a direct connection between the phenomena responsible for force generation and their effect on driving flow-induced oscillations. These tools are demonstrated in a study of the effect of cylinder shape on flow-induced vibrations. Relatively small increases in cylinder aspect-ratio are found to have a significant influence on the amplitude of oscillation, resulting in a large drop in oscillation amplitude at reduced velocities that correspond to the upper range of the synchronization regime. By mapping out the energy transfer between the fluid and structure as a function of aspect-ratio, we identify the existence of a low-amplitude stationary state as the cause of the drop in amplitude. Partitioning the fluid dynamic forces on cylinders of varying aspect-ratio then allows us to uncover the physical mechanisms behind the appearance of the underlying bifurcation. The analysis also suggests that while vortex shedding in the wake is necessary to initiate oscillations, it is the vorticity associated with the boundary layer over the cylinder that is responsible for the sustenance of flow-induced vibrations.
\end{abstract}

\begin{keywords}

\end{keywords}

\section{Introduction}
\label{sec:intro}


The flow-induced vibration of cylinders is a subject of interest in a wide variety of fields. An over-arching theme past in studies of flow-induced vibration has been the characterization of the non-linearities and associated bifurcations in the amplitude response of various systems. A case in point is the phenomenon of lock-in, which was first observed by \cite{Bishop1964TheFluid} and \cite{Feng1968TheCylinders}. The lock-in regime has been shown to include multiple response branches that exhibit subcritical behaviour \citep{Singh2005Vortex-inducedModes,Prasanth2008Vortex-inducedNumbers}, as well as a complex dependence on parameters such as Reynolds number \citep{Anagnostopoulos1992ResponseNumbers,Meneghini1995NumericalCylinder,Prasanth2008Vortex-inducedNumbers} and mass-damping \citep{Khalak1999MotionsMass-Damping}.


Ever since the first observations of these complex responses, there has been interest in identifying the underlying fluid dynamic mechanisms. These efforts have historically centered around investigations of the wake structure behind an oscillating cylinder, along with its phase relative to the oscillation. \cite{Zdravkovich1982ModificationRange.}, \cite{Ongoren1988FlowWake}, and \cite{Williamson1988VortexCylinder} were some of the first to correlate specific flow structures with the observed response of an oscillating cylinder. They showed that an abrupt jump in the phase between the lift force and oscillation was accompanied by a change in the vortex topology of the wake. \cite{Williamson1988VortexCylinder} also performed a detailed classification of the observed wake patterns as a function of oscillation kinematics, and related this to the lock-in regime. 
Motivated by these findings, there has been significant effort to relate the changes in vortex-shedding modes to the occurrence of different response branches in free-vibration \citep{Brika1993Vortex-inducedCylinder,Blackburn1999ACylinder,Govardhan2000,Carberry2005ControlledModes,Singh2005Vortex-inducedModes,Prasanth2008Vortex-inducedNumbers}. Comprehensive reviews of these efforts can be found in \cite{Williamson2004Vortex-InducedVibrations}, \cite{Sarpkaya2004AVibrations} and \cite{Bearman2011CircularVibrations}. 

These studies of the free-vibration response of cylinders have been able to provide a wealth of insight into the phenomenon of flow-induced vibrations. As outlined above, a common theme that emerges is the existence of multiple response branches, accompanied by changes in vortex shedding and/or jumps in the phase difference between the forcing and oscillation. However, the observed changes in the flow-field have only been qualitatively related to the underlying forcing mechanisms that drive the oscillations. The question of \textit{quantitatively} correlating specific flow features, such as vortex-shedding and shear layer formation, to the growth/decay of flow-induced oscillations has not been rigorously addressed. Hence the task of determining the physical mechanisms that dictate the observed complexity in the amplitude response evidently requires a two-pronged approach - quantifying the influence of specific force-producing mechanisms on the total force; and directly relating these forces to the growth/decay of flow-induced oscillations. The focus of this work is to demonstrate one such method to accomplish these tasks. 

There have been previous efforts to quantify the effects of different hydrodynamic mechanisms on the total force on an oscillating cylinder. These force decomposition methods, predominantly focused on the inertial (or added-mass effects) and viscous (or vortex) effects, have however been accompanied by vigorous debate. One such method separates potential flow (added mass) force contributions from viscous effects, or the ``vortex-flow force'' \citep{Lighthill1986FundamentalsStructures}. This was used by \cite{Govardhan2000} and \cite{Carberry2005ControlledModes} to show that different amplitude response branches are associated with changes in the vortex-shedding mode as well as the phase difference between the vortex force and the cylinder motion. However, this definition of vortex force does not explicitly separate the vorticity from other viscous contributions, and leads to the notion of ``additional vorticity'' to distinguish the inviscid vortex sheet on the cylinder from the vorticity in the rest of the flow. Furthermore, there is disagreement about the validity of isolating added-mass contributions in a general viscous flow, which assumes that this inviscid contribution operates independently from viscous effects \citep{Sarpkaya2001OnMorison}. Another commonly used force partitioning method separates the force component which is in-phase with velocity from that in-phase with acceleration \citep{Morison1950ThePiles,Sarpkaya1978FluidCylinders,Hover1998ForcesCrossflow,Gopalakrishnan1993Vortex-InducedCylinders}. While this method does not attempt to identify the force contributions from different physical mechanisms, it does isolate the in-phase component of force responsible for driving the flow-induced oscillations. It must be noted that these methods of force partitioning ignore the spatial distribution of relevant flow structures with respect to the cylinder. As a result, they do not directly relate seemingly important flow structures, such as the vortex-wake, to the forcing on the cylinder and resulting oscillation response.

In this work, we use a mathematical formulation that allows us to rigorously (from first principles) decompose the fluid dynamic forces on an immersed body into physically meaningful components. In addition, this method also allows us to quantify the effect of particular spatial regions, or flow-structures of interest, on the total force on the cylinder. The force partitioning method used here is based on the work of \cite{Quartappelle1982ForceFlows}. They showed that the forces and moments on an immersed boundary can be written as integrals of the velocity-field and its derivatives by projecting the Navier-Stokes equations onto the gradient of an auxiliary harmonic potential. This was extended by \cite{Chang1992PotentialFlow}, where the partitioning into added mass, bound- and free-vorticity were recognized. Simplifications of this method, particularly for the computation of the auxiliary potential \citep{Protas2000AnFlows,Pan2002AFlow}, and generalization for arbitrary immersed bodies and flow inhomogeneities have also been developed \citep{Howe1995OnNumbers,Magnaudet2011ANumber}. 

More recently, this force partitioning has been used to investigate the aerodynamics of insect flight by \cite{Zhang2015CentripetalInsects}, as well as the force generation mechanism of pitching airfoils \citep{Martin-Alcantara2015VortexAttack,Moriche2018OnNumber}. As shown in these studies, there are multiple advantages of such a force partitioning over those described earlier. This method allows us to separate added-mass, vorticity-induced, and shear contributions to the force on the body rigorously. Further, we will show that the ability to dissect the contribution of individual flow structures, or regions of the flow-field, is particularly valuable in the analysis of flow-induced vibrations. This method hence addresses the first of the two problems proposed above in the task of determining the physical mechanisms that dictate the flow-induced vibration of cylinders.

The second piece in determining the effect of different flow features on flow-induced vibrations, i.e. the question of quantifying the direct effect of force measurements on the observed oscillation response, has been more rigorously answered in existing literature. This has been primarily motivated by the question of using force measurements from forced vibrations to predict the free vibration response of a cylinder \citep{Sarpkaya1978FluidCylinders,Staubli1983CalculationOscillation,Gopalakrishnan1993Vortex-InducedCylinders,Hover1998ForcesCrossflow,Morse2006EmployingVibration}. Subsequently, the connection between free and forced vibration was more explicitly demonstrated in the works of \cite{Morse2009} and \cite{Kumar2016Lock-inCylinder}. They showed that the energy extracted by a cylinder undergoing forced oscillations is closely related to the amplitude response of a freely-oscillating cylinder under carefully matched kinematic conditions. This was also demonstrated in free-oscillations of airfoils in the work of \cite{Menon2019} and \cite{Zhu2020NonlinearWing}, where the computations of energy transfer using forced oscillations over a range of amplitude and frequencies (or ``energy maps'') allowed the mapping out of all possible response branches, and associated bifurcations, of a free-oscillator. This is particularly useful in systems with complicated nonlinear responses, as has been shown to be the case for oscillating cylinders \citep{Williamson2004Vortex-InducedVibrations}. Further, due to the equivalence between forced and free oscillations in terms of energy extraction, we will show here that forced oscillations and energy maps can be used to identify and isolate the parameters that cause bifurcations in order to simplify their analysis. 



The above studies hence show that force measurements can be directly related to the free-oscillation response via energy extraction. Here we use this idea, in conjunction with the partitioning of force into specific physical mechanisms and flow structures, to directly quantify the influence of each of these mechanisms on the free-oscillation response. Hence, this combination of force partitioning and energy-based analysis allows us to simultaneously address both problems outlined above - i.e. determination of the degree to which different physical mechanisms drive the flow-induced oscillations of cylinders. We show that this direct quantification of different fluid phenomena, in terms of force production and work done on the oscillating body, offers a rigorous way to assess how they interact to produce the observed bifurcations in the amplitude response. At a more fundamental level, this also uncovers insight into the mechanisms that initiate and sustain the flow-induced vibration of cylinders.


We demonstrate the efficacy of the tools described above within the context of the effect of cylinder shape on flow-induced vibrations. While the response of circular cylinders has been extensively investigated, much of the existing work on shape-effects has focused on \textit{forced} oscillations of elliptic cylinders \citep{Kanwal1955VibrationsFluid,Davidson1972JetsMotion,Hall1984OnFluid,DAlessio1999UnsteadyCylinder,DAlessio2001NumericalCylinder,Kocabiyik2004NumericalFlow}. However, changes in the aspect-ratio of the cylinder can have significant effects on the amplitude response of flow-induced vibrations \citep{Franzini2009ExperimentalCylinders,Hasheminejad2015NumericalNumbers,Wang2019Vortex-inducedFreedom}. This was shown in a comprehensive study by \cite{Navrose2014FreeNumbers}, where they identified various response branches as a function of Reynolds number for cylinders of different aspect-ratios, and mapped out the vortex shedding modes that were observed. They also showed the presence of hysteretic behaviour at the ends of the lock-in region, and jumps in phase difference between the lift force and oscillation, corresponding to changes in vortex-wake. Here, we find that changes in aspect-ratio at constant Reynolds number lead to a similarly complex behaviour. The effect of shape therefore serves as an interesting application to highlight how different force-producing mechanisms interact to drive flow-induced vibrations.  


The focus of this work is hence three-fold: (1) we demonstrate the use of a force partitioning method to dissect the contributions of different flow phenomena in driving flow-induced vibrations; (2) an energy-based analysis is performed to rigorously quantify the effect of different force-producing mechanisms on oscillating cylinders, and also to identify bifurcations in the flow-induced oscillation response of a system using energy maps; (3) an application of these tools in the context of shape-effects in flow-induced vibrations is presented to highlight the interaction of these mechanisms in producing bifurcations. 


\section{Problem description and methods}
\subsection{Computational model}
\label{sec:comp_model}

\begin{figure}
  \centerline{\includegraphics[scale=1.0]{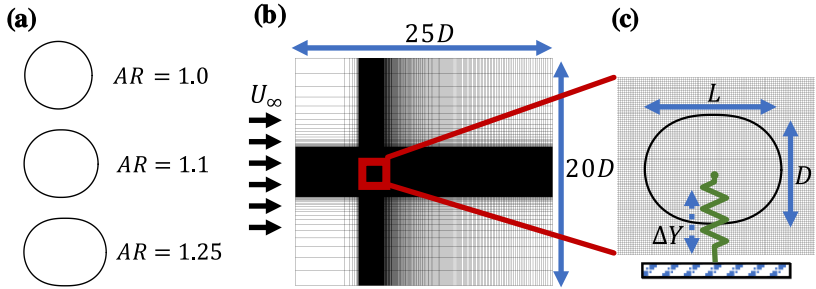}}
  \caption{Schematic of the aeroelastic system used in this study; (a) Cylinder shapes for three aspect-ratios; (b) Computational domain and grid; (c) Zoom-in of elliptic cylinder immersed in the Cartesian grid, along with dimensions and elastic model.}
\label{fig:schematic}
\end{figure}
The results reported here are obtained using two-dimensional flow simulations of the incompressible Navier-Stokes equations. The flow is coupled to elastically-mounted cylinders of different aspect-ratios that are free to oscillate transverse to the freestream. Although practical vortex-induced vibration problems can exhibit both transverse and in-line oscillations, the latter has been shown to be small in comparison to transverse oscillations \citep{Singh2005Vortex-inducedModes,Prasanth2008Vortex-inducedNumbers,Bearman2011CircularVibrations,Navrose2014FreeNumbers}, and has a negligible effect on the transverse response \citep{Williamson2004Vortex-InducedVibrations}.

This study uses a canonical model for the flow-induced vibration of cylinders, which consists of a finite-mass cylinder attached to a linear spring that allows transverse vibrations. This elastic system is immersed in a freestream flow of incompressible fluid. A schematic of this setup is shown in figure \ref{fig:schematic}. In order to study the effect of aspect-ratio, the cylinder used in this study is a generalized superellipse with its major axis aligned with the incoming flow. We use a superellipse rather than the more conventionally used ellipse in order to avoid the additional effect of increased curvature at the ends of the ellipse while varying the aspect ratio. The length and diameter of the cylinder are specified as $L$ and $D$ respectively, and the equation that governs the surface of this generalized superellipse is given by: 
\begin{equation}
    \bigg[\frac{x_s}{L/2}\bigg]^m + \bigg[\frac{y_s}{D/2}\bigg]^n = 1
    \label{eq:superellipse}
\end{equation}{}
where $x_s$ and $y_s$ are the $X$ and $Y$ coordinates of points along the surface of the cylinder. We define the aspect-ratio of the cylinder as $AR=L/D$. Further, the exponents are chosen such that $m=2L$ and $n=2$, so that this dependence of $m$ on the semi-major axes ensures that the curvature at the ends of the ellipse does not significantly increase with increasing $AR$. This is meant to approximate a ``stretched circular cylinder'' when $AR > 1$.  Throughout this work we keep $D$ fixed and vary $L$ to achieve cylinders of different aspect-ratios. The range of aspect-ratios analyzed is $1.0 < AR < 1.25$. Some representative cylinders, along with their corresponding aspect ratios, are shown in figure \ref{fig:schematic}(a). 
 
The elastic model is immersed in an incompressible fluid flow which is governed by the incompressible Navier-Stokes equations, written in dimensionless form as:
\begin{equation}
  \frac{\partial \vec{u}}{\partial t} + \vec{u}\cdot \vec{\nabla} \vec{u} = - \vec{\nabla} p + \frac{1}{Re} \vec{\nabla}^2 \vec{u} \quad ; \quad \vec{\nabla} \cdot \vec{u} = 0
  \label{eq:navier-stokes}
\end{equation}
Here the Reynolds number is given by $Re = \rho U_\infty D/\mu$, where $\rho$ is the fluid density, $U_\infty$ is the freestream velocity, $D$ is the cylinder diameter, and $\mu$ is the kinematic viscosity of the fluid. The velocity is scaled as $\vec{u} = \vec{\hat{u}}/U_\infty$, dimensionless time is given by $t = \hat{t}U_\infty/D$, and the coordinate system is scaled as $(x,y) = (\hat{x}/D,\hat{y}/D)$.

The dynamics of the elastic model is governed by the forced spring-mass system consisting of the linear spring attached to the cylinder at its center-of-mass, and a forcing given by the fluid dynamic transverse force on the cylinder, i.e., the lift force. It must be noted that there is zero structural damping in the model used here. The stiffness of the spring is given by spring constant $k$ and the mass of the cylinder is $m$. The equation for the forced spring-mass damper is non-dimensionalized using the same physical scales as for the fluid equations, $D$, $U_\infty$ and $\rho$ as the characteristic length, velocity, and density scales. Further, the dimensionless instantaneous transverse position of the cylinder's center-of-mass is denoted by $y^*$. The dimensionless equation governing the elastic system is given by:
\begin{equation}
  m^* \ddot{y}^* + k^*(y^*-y^*_{0}) = C_L
  \label{eq:spring_eq}
\end{equation}
where $m^* = 2m/(\rho D^2)$ and $k^* = 2k/(\rho U_\infty^2)$ are the dimensionless forms of the mass of the cylinder and spring stiffness. $C_L = L/(\frac{1}{2}\rho U_\infty^2 D)$ is the coefficient of lift on the cylinder (where $L$ is the lift force), and $y^*_{0}$ denotes the equilibrium position of the cylinder. We denote deflections from this equilibrium position by $\Delta y^* = (y^*-y^*_{0})$.

An important parameter in this study, and in the general literature of flow-induced vibrations, is the reduced velocity, $U^* = U_\infty/f_sD = 1/f^*_s$, where $f_s$ is the natural frequency of the spring in vacuum, and $f^*_s$ is its dimensionless form. This is related to the parameters in the above dynamical equation by $f^*_s = (1/2\pi)\sqrt{k^*/m^*}$. In this work, we study the amplitude response of the system as a function of $U^*$ by varying the spring stiffness $k^*$, while keeping $m^*$ constant at $m^*=10$. It must be noted that $m^*$ is kept constant even while the aspect-ratio of the cylinder is varied, which has the effect of isolating the effect of shape on the fluid mechanics of the problem. In other words, only the RHS of equation \ref{eq:spring_eq} depends on the aspect ratio, while the LHS is fixed for a given $U^*$. Further, $Re$ is also kept constant throughout this work, at $Re=100$. In presenting our results, dimensionless oscillation amplitude is denoted by $A^*_y$ and frequencies are reported in dimensionless form as $f^*=fD/U_\infty$.

\subsection{Numerical method}
\label{sec:num_meth}

The flow simulations in this study have been performed using the sharp-interface immersed boundary method of \cite{Mittal2008ABoundaries} and \cite{Seo2011AOscillations}. This method is particularly well-suited to fluid-structure interaction problems as it allows us the use of a simple non-conformal Cartesian grid to simulate a variety of different shapes and motions of the immersed body. Further, the ability to preserve the sharp-interface around the immersed boundary ensures very accurate computations of surface quantities. The Navier-Stokes equations are solved using a fractional-step method. Spatial derivatives are discretized using second-order central differences in space, and time-stepping is achieved using the second-order Adams-Bashforth method. The pressure Poisson equation is solved using a geometric multigrid method. This code has been extensively validated in previous studies \citep{Ghias2007,Mittal2008ABoundaries,Seo2011AOscillations}, where its ability to maintain local (near the immersed body) as well as global second-order accuracy has been demonstrated. Further, the accuracy of surface measurements has been established for a wide variety of stationary as well as moving boundary problems in these studies. For the purpose of this work, we have performed additional comparisons of results from this code with existing literature, particularly for flow-induced vibration of cylinders. Details of this are in appendix \ref{app:validation}.

The fluid-structure coupling in this work is performed sequentially, i.e., the forces on Lagrangian marker points along the surface of the cylinder are calculated at every timestep after solving the pressure Poisson equation, and the total force on the cylinder is calculated. This is then passed on to the dynamical equation for the cylinder, equation \ref{eq:spring_eq}, which is advanced in time using a second-order trapezoidal method. The cylinder is immersed in the Cartesian grid shown in figure \ref{fig:schematic}, where the size of the domain is $25D \times 20D$, and the cylinder is placed at a distance $15D$ from the downstream boundary. The grid around the cylinder is isotropic, and the grid is stretched away from the cylinder in all directions. The total grid size is $320 \times 288$ cells (see grid refinement study in Appendix \ref{app:validation}), and the resolution around the cylinder is $60$ cells across the diameter. The boundary conditions used are Dirichlet for the freestream velocity at the upstream boundary, and Neumann zero-gradient conditions at all other external boundaries.

For all cases discussed here, the cylinders are initialized at their equilibrium positions, with zero heave velocity. A uniform freestream flow, $U_{\infty}$, is specified at the upstream boundary. In the case of flow-induced oscillation, the growth of Karman vortex shedding in the wake initiates heave oscillations, which are allowed to grow until a stationary state is achieved. We refer to this maximum stationary state oscillation amplitude as $A^*_y$. For forced oscillations, the cylinder is oscillated sinusoidally with prescribed amplitude ($A^*_y$) and frequency ($f^*$) until a stationary state is achieved.

\subsection{Energy extraction}
\label{sec:energy_maps}

Here we outline a method to analyze and predict various bifurcations in the amplitude response by using the idea of ``energy maps.'' This is based on the fact that the flow-induced oscillations are driven by the energy extracted by the elastic system from the fluid flow. Over one oscillation cycle, the non-dimensional energy extracted by the heaving cylinder from the fluid, which we refer to as $E^*$, can be written as 
\begin{equation}
	E^* = \int_{t}^{t+T^*} C_L \dot{y}^*dt
	\label{eq:energy_trans}
\end{equation}
where $T^*=1/f^*$ is the dimensionless period of the cycle, and $\dot{y}^*$ is the dimensionless vertical velocity of the cylinder. It has been shown in various flow-induced oscillation systems \citep{Morse2009,Bhat2013StallNumbers,Kumar2016Lock-inCylinder,Menon2019,Zhu2020NonlinearWing} that $E^*$ can take positive as well as negative values, depending on the oscillation kinematics. In the case of flow-induced oscillations, positive values of $E^*$ lead to oscillations of increasing amplitude, while negative values lead to decaying oscillations. For a system with zero structural damping, a stationary state is reached when $E^*=0$. Hence the energy extracted by the oscillating body can be used to understand the transient as well as stationary state amplitude response of the flow-induced oscillating system.

It has also been shown that by using sinusoidal forced oscillations of the system (with prescribed kinematics) to determine $E^*$ at a range of oscillation amplitudes and frequencies, we can predict (with the assumption that the flow-induced motion is also sinusoidal) all possible response branches of a flow-induced oscillator within that range of oscillation amplitudes and frequencies \citep{Morse2009,Kumar2016Lock-inCylinder,Menon2019,Zhu2020NonlinearWing}. The values of $E^*$ predicted by forced oscillations in this domain have been shown to predict the energy extracted by flow-induced oscillations under matching operating conditions. Hence, the contours of $E^*=0$ obtained by forced oscillations define the stationary state amplitude response branches of the flow-induced oscillating system. We refer to these contours as the equilibrium curves. Furthermore, bifurcations in the equilibrium curve can be used to predict bifurcations in the amplitude response of the corresponding flow-induced oscillator. This map of energy transfer as a function of parameters of interest is referred to as an ``energy map''. In past work, this energy map has been shown to predict the stationary state amplitudes \citep{Morse2009,Kumar2016Lock-inCylinder,Menon2019,Zhu2020NonlinearWing}, transient growth \citep{Menon2019}, as well as response to perturbations \citep{menon2020aeroelastic} of a flow-induced oscillator. In this work, we will use this tool to explain the occurrence of interesting observations in the amplitude response of flow-induced oscillations.

\subsection{Force partitioning method}
\label{sec:fpm}
\newcommand{\kin}{C^{(i)}_\kappa}
\newcommand{\vif}{C^{(i)}_\omega}
\newcommand{\shr}{C^{(i)}_\sigma}
\newcommand{\pot}{C^{(i)}_\Phi}
\newcommand{\ext}{C^{(i)}_\Sigma}
\newcommand{\vol}{{V_f}}

The force partitioning method used in this study follows directly from \cite{Zhang2015CentripetalInsects} and is related to the previous work of \cite{Quartappelle1982ForceFlows} and \cite{Chang1992PotentialFlow}. An overview of this method is presented in this section, and the reader is referred to appendix \ref{app:fpm_derivation} for a detailed derivation.

\begin{figure}
  \centerline{\includegraphics[scale=1.0]{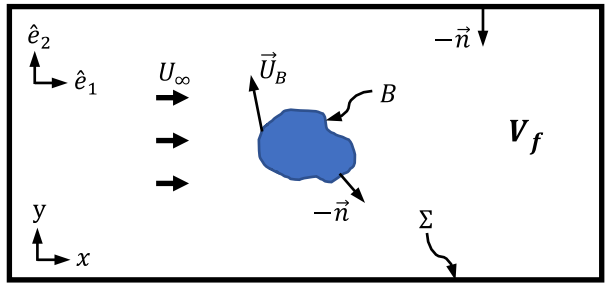}}
  \caption{Two-dimensional schematic of the domain and symbols used for the partitioning of forces on an arbitrary immersed body. Here $\vol$ is the fluid-volume in the domain, and $\Sigma$ represents the outer boundary of the domain. $B$ is the time-varying surface of the immersed body, and the unit vector pointing into the surface at every point is given by $\vec{n}$.}
\label{fig:fpm_schematic}
\end{figure}
A starting point in this method is the construction of an auxiliary potential field, given by $\phi^{(i)}$, where the superscript $(i)$ refers to the direction of force we are interested in partitioning (for instance, $i=2$ for lift and $i=1$ for drag). At any time-instance, this field depends only on the instantaneous position and shape of the immersed boundary and the outer boundary of the domain. We refer to the surface of the immersed body as $B$ and the outer boundary as $\Sigma$. A unit vector $\hat{n}$ defines the orientation at every point along these surfaces, and the fluid volume contained in the domain is referred to as $\vol$. A schematic of this setup is shown in figure \ref{fig:fpm_schematic}. Using this nomenclature, the auxiliary potential is defined as follows:
\begin{equation}
  \vec{\nabla}^2 \phi^{(i)} = 0, \ \ \mathrm{ with } \ \
  \vec{n} \cdot \vec{\nabla} \phi^{(i)}=
    \begin{cases}
      n_i \;, \; \mathrm{on} \; B \\
      0 \; \;, \; \mathrm{on} \; \Sigma \\
    \end{cases}
  \label{eq:scalar}
\end{equation}


The partitioning of forces on the body, $B$, is then obtained from projecting the Navier-Stokes equation on the field $\nabla \phi^{(i)}$, and performing a volume integral over the fluid domain $\vol$. We present this partitioning in the form of non-dimensional force coefficients below, where $C_i = F_i/(\frac{1}{2}\rho U_{\infty}^2D)$ is the coefficient of force on the body in the $i$-direction (and $F_i$ is the dimensional form of this force). This force can be partitioned into components corresponding to different force-producing mechanisms as:
\begin{equation}
  C_i = \kin + \vif + \shr + \pot + \ext
  \label{eq:force_decomp_initial}
\end{equation}
Here $\kin$ is the kinematic force, $\vif$ is the vorticity-induced force, $\shr$ is the force due to viscous effects, $\pot$ is the force associated with the corresponding potential flow-field, and $\ext$ is related to effects of the outer boundary. These force components take the following form:
\begin{align}
    \kin &= - \int_B \vec{n} \cdot \frac{d\vec{U}_B}{dt}\phi^{(i)} dS - \int_B \frac{1}{2}|\vec{U}_B|^2 n_i dS \label{eq:kin}\\
    \vif &=  \int_\vol \bigg\{\Big[\vec{\nabla} \cdot (\vec{\omega} \times \vec{u}) \Big] \phi^{(i)} + \vec{\nabla} \cdot \Big[ \vec{\nabla} \Big( \frac{1}{2}\vec{u}_v \cdot \vec{u}_v  + \vec{u}_\Phi \cdot \vec{u}_v \Big) \phi^{(i)}\Big] \bigg\} dV \label{eq:vif}\\
    \shr &= \frac{1}{Re} \int_B (\vec{\omega} \times \vec{n}) \cdot \bigg(\vec{\nabla} \phi^{(i)}-\hat{e}_i\bigg) dS \label{eq:shr} \\
    \pot &= \int_\vol \vec{\nabla} \cdot \bigg[ \vec{\nabla}\bigg( \frac{1}{2}\vec{u}_{\Phi} \cdot \vec{u}_{\Phi} \bigg) \; \phi^{(i)} \bigg] dV \label{eq:pot}\\
    \ext &= \int_\Sigma \bigg\{  - \vec{n} \cdot \frac{d\vec{u}}{dt}\phi^{(i)} + \frac{1}{Re} (\vec{\omega} \times \vec{n}) \cdot \vec{\nabla} \phi^{(i)} \bigg\} dS. \label{eq:ext}
\end{align}
Here $\vec{U}_B$ is the time-varying velocity of every point along the immersed-boundary surface, and $\hat{e}_i$ is the standard basis vector in the $i$-direction. These quantities are shown in figure \ref{fig:fpm_schematic}. The velocity components, $\vec{u}_v$ and $\vec{u}_\Phi$, are the vorticity-related and curl-free (potential flow) components of the flow, respectively. These are obtained via a Helmholtz decomposition of the velocity field \citep{BATCHELOR} (see equation \ref{eq:helmholtz_deriv}). 

We see from equation \ref{eq:kin} that the kinematic force, $\kin$, is purely determined by the velocity ($\vec{U}_B$) and shape ($B$; $\vec{n}$) of the immersed surface. Further, one can show that the first term in equation \ref{eq:kin} is related to the added-mass effects, and the second term is associated with the centripetal acceleration reaction on the immersed body (see  \cite{Zhang2015MechanismsInsects} for details). Since there is no body rotation in the current case, this second term is identically equal to zero.

The vorticity-induced force, $\vif$ in equation \ref{eq:vif}, is a volume integral over the fluid domain that depends on the vorticity as well as the flow-field associated with the rotational velocity $\vec{u}_v$. It is important to point out that this force is zero in the absence of vorticity (because $\vec{u}_v=0$ in that case), and hence characterizes the force production from purely vorticity-dependent effects. Further, the fact that this term is a volume integral allows us to separate the effect of individual vortices, or the vorticity in various parts of the flow field, which we show is very insightful in this work. 

The viscous term, $\shr$, has two components and consists of the effects from the pressure force associated with momentum diffusion at the wall as well as the direct contribution from the wall shear. The remaining two terms, the force associated with purely irrotational effects, $\pot$, and the force due to the flow and vorticity at the outer boundary, $\ext$, can be shown to go to zero when the domain is sufficiently large \citep{Zhang2015MechanismsInsects}.

\begin{figure}
  \centerline{\includegraphics[scale=1.0]{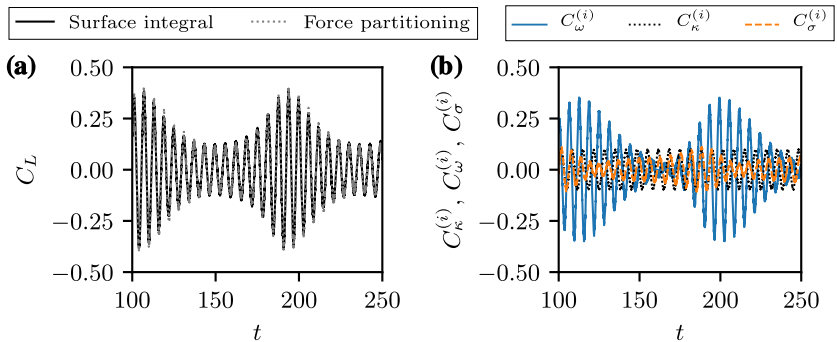}}
  \caption{Partitioning of the lift force on a cylinder with aspect-ratio $AR=1.15$, oscillation amplitude $A^*_y=0.05$, and oscillation frequency $f^*=0.15$. (a) Total lift-force compared with sum of force contributions from vorticity-induced ($\vif$), kinematic ($\kin$), and viscous force ($\shr$) components; (b) Time-series plots of vorticity-induced, kinematic, and viscous force contributions to lift.}
\label{fig:fpm_val}
\end{figure}
We now show a sample force partitioning for a forced-oscillation case with $AR=1.15$ and $A^*_y=0.05$, to highlight the most important components of the total force. In figure \ref{fig:fpm_val} we focus on the contributions from three components -- the vorticity-induced force, kinematic force and viscous force to the lift force on the cylinder. A comparison of the sum of contributions from these terms $(\vif + \kin + \shr)$ with the total lift force on the cylinder is shown in figure \ref{fig:fpm_val}(a). We see that these components account for most of the force on the oscillating cylinder. This is expected for domains that are sufficiently large \citep{Zhang2015MechanismsInsects}. This also verifies the adequacy of the integration domain used here. In figure \ref{fig:fpm_val}(b) we show time-series plots of the vorticity-induced force, kinematic force, and viscous force for this case. We see that $\kin$ shows a sinusoidal behaviour, which is dictated by the sinusoidal oscillation of the cylinder due to the fact that it is related to added-mass effects in the absence of rotation \citep{Zhang2015MechanismsInsects}. Another observation from figure \ref{fig:fpm_val}(b) is that the vorticity-induced force accounts for the bulk of the force on the cylinder during periods of high lift force. The amplitude of viscous force oscillations too shows variations, however the maximum amplitude is much smaller than that of $\vif$.

Henceforth this work will focus only on the effects of the kinematic force, the vorticity-induced force, and the viscous force. Further, we are exclusively interested in the partitioning of the lift force coefficient, $C_L$. We will therefore dispense with the superscript $i$ and simply use the following convention
\renewcommand{\kin}{C_\kappa}
\renewcommand{\vif}{C_\omega}
\renewcommand{\shr}{C_\sigma}
\renewcommand{\pot}{C_\Phi}
\renewcommand{\ext}{C_\Sigma}
\begin{equation}
  C_L = \kin + \vif + \shr + \pot + \ext
  \approx \kin + \vif + \shr 
  \label{eq:force_decomp}
\end{equation}
Combining this with calculations of the energy extraction discussed in section \ref{sec:energy_maps}, the total energy gained by the cylinder in each oscillation cycle can correspondingly be partitioned into contributions from each component in the force partitioning using equation \ref{eq:energy_trans}. This is given by:
\begin{equation}
 E^* \approx E^*_\kappa + E^*_\omega + E^*_\sigma 
  \label{eq:energy_decomp}
\end{equation}

\section{Aspect-ratio effects in flow-induced vibrations}
\label{sec:flow_induced_energy}

We begin our discussion of the results by briefly describing how changes in the the amplitude response and energy extracted by the cylinder from the flow are affected by the aspect-ratio of the oscillating cylinder. This discussion serves to highlight non-linearities in the amplitude response, dictated by the fact that changes in shape lead to changes in the relative contributions of the underlying mechanisms that drive the oscillation. The behaviour discussed here will set the context for the subsequent application of the aforementioned partitioning methods in analyzing complex oscillation responses, and uncovering the origin of the bifurcations in the amplitude response. 




\subsection{Flow-induced oscillation response}
\label{sec:flow_induced}
\begin{figure}
  \centerline{\includegraphics[scale=1.0]{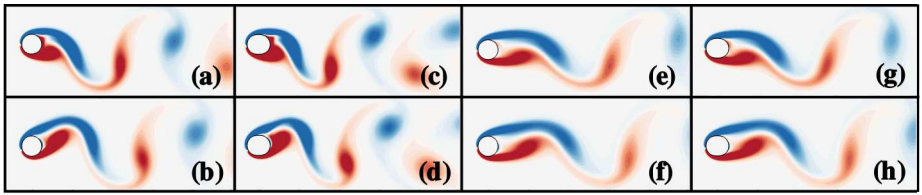}}
  \caption{ Snapshots of the flow around transversely oscillating cylinders with different aspect ratios ($AR$) and reduced velocity ($U^*$), at $Re=100$. The flow is visualized using contours of $Z$-vorticity. The top panel shows snapshots close to the upper maximum of the oscillation cycle for each case, and the bottom panel show snapshots close to the mean position as the cylinder is moving downwards. Cases shown in (a)-(d) show large amplitude oscillations (of similar magnitude, $A^*_y \approx 0.49$) and (e)-(h) show relatively smaller oscillation amplitudes ($A^*_y \approx 0.28$); (a)-(b) $AR = 1.0$, $U^* = 6.0$; (c)-(d) $AR = 1.15$, $U^* = 5.0$; (e)-(f) $AR = 1.0$, $U^* = 7.5$; (c)-(d) $AR = 1.15$, $U^* = 7.0$.  }
\label{fig:snapshots}
\end{figure}
We begin with a qualitative description of the flow-fields for some select flow-induced oscillation cases. Figure \ref{fig:snapshots} shows instantaneous vorticity contours of the flow for four different flow-induced oscillation cases at $Re=100$, with aspect ratios $AR=1.0$ and $AR=1.15$, and reduced velocity varying from $U^*=5.0$ to $U^*=7.5$. The snapshots are shown after each system has achieved stationary state oscillations. The top panel in figure \ref{fig:snapshots} shows snapshots of the flow at the time instance when the cylinder is at the position of maximum displacement, and the bottom panel shows snapshots when the cylinder is close to the mean position during the downward motion (position of maximum velocity). Further, figures \ref{fig:snapshots}(a)-(b) and \ref{fig:snapshots}(c)-(d) show two cases exhibiting large amplitude oscillations, with $A^*_y \approx 0.49$, while the two cases in figures \ref{fig:snapshots}(e)-(f) and \ref{fig:snapshots}(g)-(h) exhibit smaller amplitudes, with $A^*_y \approx 0.28$.

It is immediately apparent that the cases with larger amplitude oscillations, in figures \ref{fig:snapshots}(a)-(d), have wake vortices that are more closely spaced than the smaller amplitude cases in figures \ref{fig:snapshots}(e)-(h). This can be qualitatively assessed by simply counting the number of vortices within the frame of the flow visualizations shown. The reason for this is presumably because larger oscillation amplitudes lead to stronger shear layers, and thus earlier roll-up. Further, even between the two cases with large amplitude, the case with $AR=1.0$ and $U^* = 6.0$, in figures \ref{fig:snapshots}(a)-(b), shows a less tight (and possibly less staggered) vortex pattern in the wake as compared to the case with $AR = 1.15$ and $U^* = 5.0$, which is shown in figures \ref{fig:snapshots}(c)-(d). At the smaller amplitude of oscillation, the difference in wake shedding pattern between the two cases shown in figures \ref{fig:snapshots}(e)-(f) and \ref{fig:snapshots}(g)-(h) is less evident. However, we still see that the case with $AR=1.0$ and $U^*=7.5$, in figures \ref{fig:snapshots}(e)-(f), shows a slightly larger spatial wavelength in the wake (in the streamwise direction) than the case with $AR=1.15$ and $U^*=7.0$, in figures \ref{fig:snapshots}(g)-(h). Hence we see that, although the two large amplitude cases have similar oscillation amplitudes, as do the two smaller amplitude cases, a modest change in shape results in a qualitatively different flow. However, these plots and corresponding data do not immediately reveal the role that changes in vortex shedding patterns have on the flow-induced oscillation.

\begin{figure}
  \centerline{\includegraphics[scale=1.0]{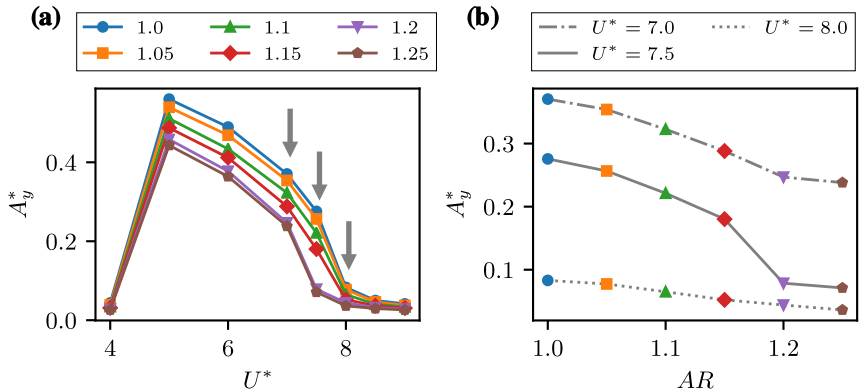}}
  \caption{ Heave amplitude response of flow-induced oscillations at $Re=100$; (a) Maximum heave amplitude ($A^*_y$) versus reduced velocity ($U^*$), plotted for cylinders of various aspect ratios ($AR$); (b) Maximum heave amplitude as a function aspect ratio at three relevant values of $U^*$, i.e. $U^* = 7.0,7.5,8.0$. Symbols for different aspect ratios correspond to those in (a). }
\label{fig:heave_re100}
\end{figure}
We now discuss the stationary-state amplitude response of flow-induced oscillations for a range of $U^*$ and $AR$. In figure \ref{fig:heave_re100}(a) we plot the maximum heave amplitude ($A^*_y$) at $Re=100$ as a function of $U^*$ for cylinders of different aspect-ratios. We see that the amplitude curves for all the aspect-ratios studied here show the familiar peak that is associated with the initial and lower branch of the cylinder response \citep{Williamson2004Vortex-InducedVibrations}. Further, the smaller aspect ratios show larger oscillation amplitudes through the entire range of $U^*$ investigated here. An interesting observation from figure \ref{fig:heave_re100} is that the $U^*$-extent of the synchronization regime depends on the aspect-ratio. This is particularly evident at the higher-$U^*$ limit of the synchronization regime ($U^* \approx 7.5$), where we see that higher $AR$ cylinders have a smaller synchronization region. 

To more clearly underscore the effect of this $AR$-dependence on the amplitude response, figure \ref{fig:heave_re100}(b) shows the heave amplitude response as a function of cylinder aspect-ratio for three select values of $U^*$. These values are indicated by arrows in figure \ref{fig:heave_re100}(a), i.e. $U^* = 7.0$, $7.5$ and $U^*=8.0$. At $U^*=7.0$, we see that the cylinders all show large amplitude oscillations, and the response is within the lower branch for all values of $AR$. At $U^*=8.0$, which lies beyond the lower branch (the so-called desynchronization regime) for all values of $AR$, we see that $A^*_y$ is small and decreases approximately linearly with increasing $AR$. In-between these values of $U^*$, i.e. at $U^*=7.5$, the smaller aspect-ratio cylinders seem to be on the lower branch whereas the cylinders with $AR \geq 1.2$ show small amplitude oscillations with amplitudes close to those seen at $U^*=8.0$, which is in the desynchronization regime. There is evidently a sharp transition between the large- and small-amplitude response (between the behaviour seen at $U^* = 7.0$ and $U^* = 8.0$) when the aspect-ratio is increased beyond $AR \geq 1.2$ at $U^* \approx 7.5$.  In particular, there is a $2.3\times$ drop in amplitude for 4\% increase in $AR$ at $U^*=7.5$, compared with an approximately $1.2\times$ drop in amplitude for $U^*=7.0$ and $U^*=8.0$. It is interesting to note that this transition becomes more abrupt with increasing Reynolds number (although at different $U^*$ and $AR$ values), as shown in appendix \ref{app:re250} for $Re=250$.


\subsection{Energy transfer}
\label{sec:energy_trans}
In order to utilize the force and energy partitioning tools outlined previously to analyze the reasons behind this transition in amplitude, it is essential to first identify the bifurcation causing this behaviour, and the parameters it depends on. As mentioned earlier in section \ref{sec:energy_maps}, energy maps allow us to identify all possible response branches of a flow-induced oscillator, along with their associated bifurcations, by using sinusoidal forced oscillations at matching operating conditions. In the present work, we are interested in bifurcations in the amplitude response as a function of not just the amplitude and frequency of oscillation (as has been done in previous works), but also the cylinder aspect-ratio. 

The addition of shape as a parameter in energy transfer would require the generation of three-dimensional energy maps in order to achieve matching conditions in frequency, amplitude, and shape, and this is computationally expensive. To circumvent this issue, we choose to probe the three-dimensional energy landscape of the system at the oscillation frequency expected for each cylinder, which is its observed flow-induced oscillation frequency. While this frequency is mostly determined by the structural parameters, added-mass effects (which are different for cylinders of different shapes) can generate small variations in this frequency as per $f^* = (1/2 \pi) \sqrt{k/\left(m+m_{a} \right) }$, where $m_a$ denotes added-mass. Hence for given structural parameters, the observed flow-induced oscillation frequency varies as a function of shape. In the current analysis we focus on the behaviour corresponding to $U^*=7.5$, as it is this value of $U^*$ at which we observe the abrupt drop in flow-induced oscillation amplitude discussed in the previous section ($\S$ \ref{sec:flow_induced}). For this case, $f^*$ is determined via the corresponding flow-induced oscillation for each cylinder, at $U^*=7.5$, and is found to range from  $0.13 \lessapprox f^* \lessapprox 0.155$. As mentioned above, this variation is due to added-mass effects associated with varying aspect-ratio, and also the fact that the smaller-$AR$ cylinders are within the synchronization regime at $U^*=7.5$, whereas the larger-$AR$ cylinders are not (as seen in figure \ref{fig:heave_re100}). 

In order to map out the amplitude response branches as a function of shape at $U^*=7.5$, we perform forced oscillation simulations where each cylinder with a given aspect-ratio is forced to oscillate sinusoidally with a frequency corresponding to its dominant stationary state flow-induced oscillation frequency that is observed at $U^*=7.5$. Further, we prescribe oscillation amplitudes in the range $0 <A^*_y \leq 0.5$ for each cylinder, and perform approximately 100 such simulations for the set of cylinders. The simulations are allowed to reach a stationary state, and the energy transfer, $E^*$ given by equation \ref{eq:energy_trans}, is then computed for each case and analyzed as a function of oscillation amplitude and cylinder aspect-ratio.

\begin{figure}
  \centerline{\includegraphics[scale=1.0]{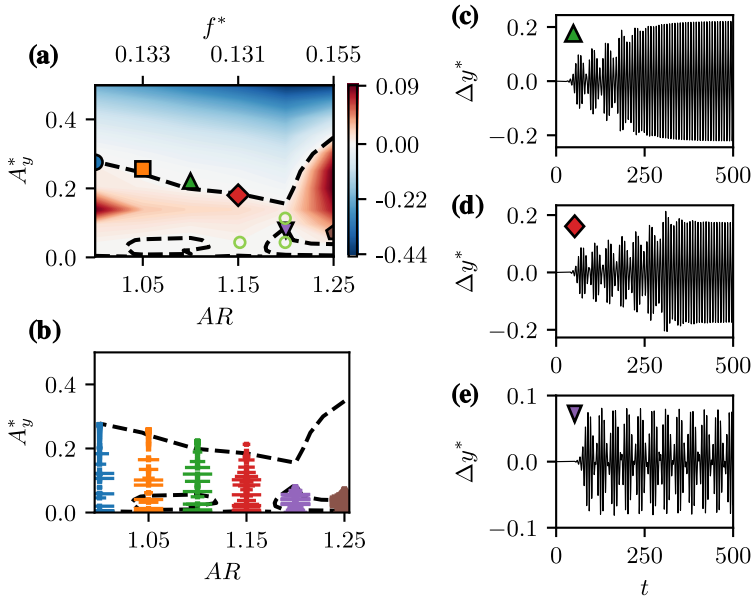}}
  \caption{(a) Contours of energy transfer between the cylinder and flow, as a function of oscillation amplitude ($A^*_y$) and aspect ratio ($AR$). This energy transfer is computed using forced oscillations. The dashed lines show the zero energy transfer contours. Stationary state amplitudes of flow-induced oscillation for cylinders of different aspect ratios, at $U^*=7.5$, are plotted using symbols as in figure \ref{fig:heave_re100}. Light-green circles show $A^*_y$ and $AR$ of the forced-oscillation cases analyzed in section \ref{sec:fpm_results}; (b) Trajectory of oscillation amplitude during each cycle for the flow-induced oscillation cases at $U^*=7.5$. The horizontal lines indicate the magnitude of peak-to-peak amplitude difference between successive cycles, indicating non-sinusoidal behaviour}; (c)-(e) Flow-induced oscillation response for cases with $U^*=7.5$, and aspect ratios $AR=1.1$, $AR=1.15$ and $AR=1.2$ respectively.
\label{fig:energy_maps}
\end{figure}
In figure \ref{fig:energy_maps}(a), we show a contour plot of the energy extracted by the oscillating cylinder (undergoing forced oscillations) as a function of heave amplitude $A^*_y$, and cylinder aspect-ratio $AR$. The dashed lines show the contours of zero energy transfer, or the equilibrium curves. These equilibrium curves represent the stationary-states of sinusoidal flow-induced oscillators with aspect-ratios $1.0 \leq AR \leq 1.25$ and frequencies corresponding to $U^*=7.5$.  These energy map predictions of the flow-induced oscillation response branches are verified by superimposing the observed stationary state oscillation amplitudes of the flow-induced oscillation cases at $U^*=7.5$ on the energy map, using the same symbols as in figure \ref{fig:heave_re100}. We see that the stationary state amplitudes do in fact all lie along or near the equilibrium curves in figure \ref{fig:energy_maps}(a).  

It must however be noted that the flow-induced oscillations at $U^*=7.5$ do not necessarily exhibit pure sinusoidal behaviour, especially at small oscillation amplitudes, where natural vortex shedding continues to play an important role. Three examples of this non-sinusoidal behaviour are shown in figures \ref{fig:energy_maps}(c)-\ref{fig:energy_maps}(e), for cases with $AR=1.1$, $AR=1.15$  and $AR=1.2$ respectively. The presence of a beat phenomenon is clearly observed in these cases, where the oscillation is modulated by the vortex shedding frequency. The degree of non-sinusoidal behaviour during the progression of the cylinder oscillation is shown in figure \ref{fig:energy_maps}(b). This figure shows the trajectory of amplitude growth, from $A^*_y=0$ to the stationary state amplitude, for each flow-induced oscillation case at $U^*=7.5$. Along this amplitude trajectory, the size of the horizontal lines (similar to error bars) indicate the magnitude of the amplitude difference between successive cycles, and are therefore a direct measure of departure from sinusoidal behavior. The highly non-sinusoidal behaviour at small amplitudes is clear from the larger horizontal bars at small amplitudes for all cases. The length of these bars progressively gets shorter at larger amplitudes, for $AR \leq 1.15$, indicating the emergence of sinusoidal behaviour as the oscillation amplitude increases. Hence, although the equilibrium curves are indicative of the expected flow-induced oscillation behaviour, we do not expect them to exactly match all the exhibited behavior. Indeed, the correspondence between the equilibrium curves of the energy map and the stationary state amplitude is not exact, especially for cases with small amplitudes.

An interesting observation from the energy map is the presence of multiple equilibria for a given shape - one at large amplitude that stretches across all aspect-ratios, which we will refer to as the high-amplitude equilibrium curve; and another set of curves at small amplitudes. However, amongst the equilibrium curves at small amplitudes, we find that the one at aspect-ratios $1.05 \lessapprox AR \lessapprox 1.10$ is a significantly weaker, and less stable equilibrium than that at $AR \gtrapprox 1.20$. As shown in \cite{Menon2019}, the stability of an equilibrium curve on the energy map is given by the sign of $dE^*/dA^*_y$. The conditions for a stable equilibrium are $E^*=0$ and $dE^*/dA^*_y<0$. Comparing the energy extraction and gradients enclosed by the two equilibrium curves at small amplitudes, we find that (not shown here for brevity) the region at aspect-ratios $AR \gtrapprox 1.20$ has roughly $2\times$ the energy loss, and energy gradients that are roughly an order of magnitude more negative, than that at $1.05 \lessapprox AR \lessapprox 1.10$. This large difference in stability is especially important due to the presence of beating oscillations at small amplitudes, as shown in figure \ref{fig:energy_maps}(b). The fact that beating oscillations can be written as sum of sine waves with different frequencies suggests that the presence of beats in the flow-induced oscillation can be considered as a ``frequency perturbation'' from the single-frequency sinusoidal oscillation at which the energy extraction is computed for each aspect-ratio. Hence the strength and stability of these equilibria is an important consideration in view of this perturbation from the energy map's requirement of matching kinematics between the forced and flow-induced oscillations. This analysis, relating to the robustness of the two low-amplitude equilibrium curves and the presence of beats in the flow-induced oscillation, is examined in more detail in appendix \ref{app:energy_map_15}. There we show that the equilibrium curve at $AR \gtrapprox 1.20$ persists when the energy map is computed at a slightly different frequency, which corresponds to the second peak in the frequency spectrum of the beating oscillations. However, the equilibrium curve at $1.05 \lessapprox AR \lessapprox 1.10$ is not present at this secondary frequency. Due to these reasons, as we will show below, the former plays a larger role in the dynamics, and will be referred to as the low-amplitude equilibrium curve.



The topology of the energy map, and the presence of multiple equilibria for some aspect-ratios, suggests an explanation for the observed bifurcation in the amplitude response at $U^*=7.5$. For values of $AR$ that have multiple equilibrium curves at different oscillation amplitudes, we expect a system with an initial condition of zero amplitude (as is the case in this study) to reach its stationary state at the lowest-amplitude equilibrium corresponding to its value of $AR$. This is due to the fact that the amplitude of oscillation in these cases grows from $A^*_y=0$ (starting at the bottom of the energy map) at constant $AR$, until the growth is stopped on the first encounter with a stable equilibrium. Hence for the energy map in figure \ref{fig:energy_maps}(a), the first stationary state encountered by cylinders with $AR \gtrapprox 1.15$ is the low-amplitude equilibrium curve. On the other hand, due to the fact that the low-amplitude equilibrium curve at aspect-ratios $1.05 \lessapprox AR \lessapprox 1.10$ is a weak equilibrium, as explained above and in appendix \ref{app:energy_map_15}, and the oscillations at these amplitudes are non-sinusoidal, the low-amplitude equilibrium curve at $1.05 \lessapprox AR \lessapprox 1.10$ does not stop the amplitude growth of cylinders with these aspect-ratios. As a result, cylinders with $AR \lessapprox 1.15$ are able to grow to larger amplitudes and settle on the high-amplitude equilibrium curve. Therefore there is a bifurcation from the large-amplitude branch to the small amplitude branch, which occurs at $AR \gtrapprox 1.15$. Hence we see that the mapping out of all possible response branches of the system allows a better understanding of the bifurcation observed in the flow-induced oscillation response.

We see from the above discussion that the abrupt drop in oscillation amplitude seen in the case of $U^*=7.5$ is due to a bifurcation associated with the emergence of a low-amplitude response branch for $AR \gtrapprox 1.15$. Moreover, we have found (see appendix \ref{app:energy_map_15}) that this low-amplitude branch occurs even without the minor variations in $f^*$ for each aspect-ratio. Consequently, the bifurcation appears to be caused purely by the change in cylinder shape. This suggests that there is a change in the forcing mechanism(s) on the cylinder as we increase the cylinder aspect-ratio for low-amplitude oscillations. It must be pointed out that the ability to identify the nature of the bifurcation, and isolate the parameters related to its occurrence is a particular merit of the energy map approach in this context. In what follows, we attempt to dissect the physical origin of this bifurcation, and explain the interaction of forcing mechanisms that give rise to its emergence.

\section{The mechanisms that govern flow-induced vibrations}
\label{sec:fpm_results}

We focus now on the main theme of this work - which is to demonstrate the use of the force partitioning method, in conjunction with an energy-based analysis, to analyze the fluid-dynamic mechanisms that drive flow-induced vibration of cylinders. The aim will be to highlight the relative contributions of different physical mechanisms in the initiation and sustenance of oscillations, as well as to highlight the use of these tools in analyzing bifurcations such as those described in $\S$ \ref{sec:flow_induced_energy}. In past studies, such bifurcations have been related to changes in vortex-shedding patterns, however the direct influence of these effects on the forces on the cylinder has been mostly analyzed qualitatively. In this work, by quantifying the energy extracted by the oscillating cylinder from different physical mechanisms, we establish a rigorous way to connect distinct flow mechanisms with the emergence of oscillations. In order to uncover the origin of the bifurcation seen in section \ref{sec:flow_induced_energy}, we analyze the flow physics and forcing mechanisms for some select forced-oscillation cases around the low-amplitude equilibrium curve discussed earlier. Through this, we show that we are able to develop insights that can then be tied back to the fundamental mechanisms driving flow-induced oscillations.

\begin{figure}
  \centerline{\includegraphics[scale=1.0]{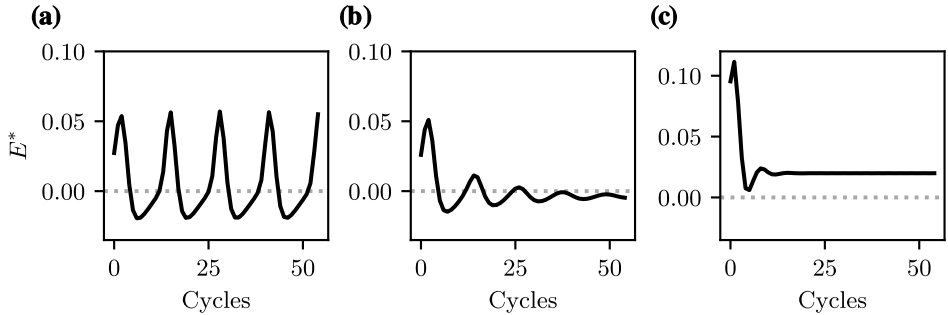}}
  \caption{Energy extraction ($E^*$) during each oscillation cycle for cylinders of different aspect ratios and oscillation amplitudes at $Re=100$ and $f^*=0.15$; (a) $AR = 1.15$, $A^*_y=0.05$; (b) $AR = 1.20$, $A^*_y=0.05$; (c) $AR = 1.20$, $A^*_y=0.10$.}
\label{fig:energy_timeseries}
\end{figure}
In this discussion, we analyze three forced-oscillation cases close to the bifurcation of interest, keeping the frequency constant at $f^*=0.15$ for simplicity. This frequency is very close to that of the flow-induced oscillation cases that are involved in the bifurcation. The aspect-ratios and oscillation amplitudes of the cases analyzed here are shown (using light-green circles) on the energy map in figure \ref{fig:energy_maps}(a). The effect of $AR$ is studied by comparing two cases with $AR = 1.15$ and $AR=1.2$ and constant amplitude of $A^*_y=0.05$, and the effect of $A^*_y$ is studied by comparing $A^*_y=0.05$ with $A^*_y=0.10$ for cases with $AR=1.2$. 

Time-series plots of energy extraction during each oscillation cycle for these cases are shown in figures \ref{fig:energy_timeseries}(a)-(c). We see that the energy extraction in the case with $AR=1.15$ and $A^*_y=0.05$, shown in figure \ref{fig:energy_timeseries}(a), exhibits a well-defined oscillation in time. This indicates that the oscillation frequency of $C_L$ is not equal to the heave oscillation frequency ($f^*=0.15$), i.e. the vortex shedding is not locked-in. The mean energy extraction is positive in this case, with $E^*=7.12 \times 10^{-3}$. The energy transfer in the case with $AR=1.2$ and $A^*_y=0.05$, plotted in figure \ref{fig:energy_timeseries}(b), shows initial oscillations but eventually settles down to a constant value. This final stationary state, with $E^*=-2.92 \times 10^{-3}$, is within the island of $E^*<0$ corresponding to the low-amplitude branch of the equilibrium curve. Finally, the larger-amplitude case with $AR=1.2$ and  $A^*_y=0.10$, shown in figure \ref{fig:energy_timeseries}(c), also shows a constant value of $E^*$ at stationary state, which is positive and much larger than the other two cases, at $E^*=2.46 \times 10^{-2}$. Further, the locked-in stationary state is achieved much faster in the case of $A^*_y=0.10$ than in the case of $A^*_y=0.05$ due to the fact that larger oscillation amplitudes drive the forcing frequency to a locked-in state.

\subsection{Vorticity and viscous effects in sustained vibration}

We now analyze the partitioning of the energy extraction for the cases discussed above, using the force partitioning method described in section \ref{sec:fpm}. As discussed in section \ref{sec:fpm}, this method allows us to partition the total force on the oscillating cylinder into physically relevant components, shown in equation \ref{eq:force_decomp}. A key point in the analysis presented here is that each of these force components does work on the cylinder, contributing to the total energy extracted by the oscillating cylinder. We can therefore use this force partitioning to compute the energy delivered to the oscillating cylinder by each of these forcing mechanisms as in  \ref{eq:energy_decomp}. This allows us to quantify the effect of these force-producing mechanisms in driving the oscillations. In our discussion here, we will focus only on the force/energy contributions from the kinematic force ($\kin$), vorticity-induced force ($\vif$), and viscosity-related ($\shr$) components in the partitioning. This is because, as was verified in section \ref{sec:fpm} (figure \ref{fig:fpm_val}), these components account for nearly all the force on the cylinder.

\begin{figure}
  \centerline{\includegraphics[scale=1.0]{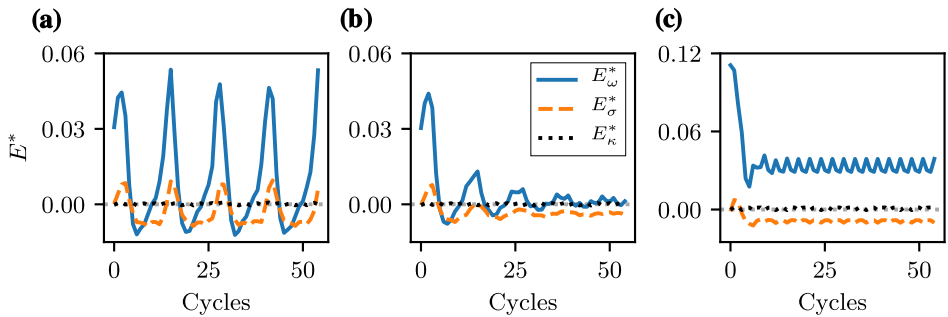}}
  \caption{ Partitioning of the energy extraction per oscillation cycle into contributions from vorticity-induced force (blue, solid line), viscous force (orange, dashed line), and kinematic force (black, dotted line). This is shown for cylinders of different aspect ratios, oscillating with different amplitudes; (a) $AR = 1.15$, $A^*_y=0.05$; (b) $AR = 1.20$, $A^*_y=0.05$; (c) $AR = 1.20$, $A^*_y=0.10$.}
\label{fig:energy_fpm}
\end{figure}
In figures \ref{fig:energy_fpm}(a)-(c) we show the energy extracted by the oscillating cylinder from added-mass effects ($E^*_\kappa$), viscous contributions ($E^*_{\sigma}$), and vorticity-related contributions ($E^*_{\omega}$) for the three cases mentioned above. In all three cases we see that the bulk of the energy extraction is driven by the vorticity-induced force. Another observation that is common to the three cases is that the kinematic force contributes nearly zero net energy transfer. This is expected for periodic, non-rotating oscillations of the cylinder. It is easy to show that the second term on the right-hand side of equation \ref{eq:kin} is zero in the absence of rotation, and the first term corresponds to the traditional added-mass force in ideal fluids \citep{Zhang2015MechanismsInsects}. Hence by plugging this first term, which is a sinusoidal term in-phase with the displacement of the cylinder, into the equation for energy transfer (equation \ref{eq:energy_trans}), we see that this integral is zero due to the periodicity of the system. Therefore, the total energy transfer is primarily determined by the contributions from the vorticity-induced force and viscous effects. 

\begin{table}
\centering
\begin{tabular}{||c c c c c c c||} 
 \hline
 $A^*_y$ & $AR$ & $E^*$ & $E^*_\omega$ & $E^*_\sigma$ & $E^*_S$ & $E^*_W$ \\ [0.5ex] 
 \hline\hline
 0.05 & 1.15 & $7.12 \times 10^{-3}$ & $9.98 \times 10^{-3}$ & $-3.02 \times 10^{-3}$ & $1.55 \times 10^{-2}$ & $-1.78 \times 10^{-3}$ \\ 
 0.05 & 1.20 & $-2.92 \times 10^{-3}$ & $7.50 \times 10^{-4}$ & $-3.73 \times 10^{-3}$ & $1.05 \times 10^{-2}$ & $-5.67 \times 10^{-3}$ \\
 0.10 & 1.20 & $2.46 \times 10^{-2}$ & $3.27 \times 10^{-2}$ & $-8.87 \times 10^{-3}$ & $6.37 \times 10^{-2}$ & $-1.25 \times 10^{-2}$ \\ [0.5ex] 
 \hline
\end{tabular}
\caption{Summary of mean energy extraction (total energy and components of the energy partitioning) for three forced-oscillation cases with different aspect-ratios and oscillation amplitudes.}
\label{table:energy_mean}
\end{table}
In the case with $AR=1.15$ and $A^*_y=0.05$, shown in figure \ref{fig:energy_fpm}(a), the mean energy extracted by $\vif$ is $E^*_\omega = 9.98 \times 10^{-3}$ and that from $\shr$ is $E^*_\sigma = -3.02 \times 10^{-3}$. As was seen in the case of the total energy, the individual contributions to the total energy also show oscillations in time due to the fact that the forcing is not locked-in. The case with $AR=1.2$ and $A^*_y=0.05$ is shown in figure \ref{fig:energy_fpm}(b) and we see that energy extraction from the vorticity is very small in this case, with $E^*_\omega = 7.50 \times 10^{-4}$. The negative energy transfer from viscous effects, $E^*_\sigma = -3.73 \times 10^{-3}$, pushes the total energy extraction below zero. In the high-amplitude case with $AR=1.2$ and $A^*_y=0.10$, shown in figure \ref{fig:energy_fpm}(c), we see that the energy extracted from the vorticity-induced force has a mean positive value of $E^*_\omega = 3.27 \times 10^{-2}$, and this is much larger than in the other two cases. This case too shows energy loss due to the viscosity-related effects, $E^*_\sigma = -8.87 \times 10^{-3}$, however the large value of $E^*_\omega$ compared to the previous case ensures that the total energy extraction is positive in this case. A summary of these mean energy extraction values is shown in table \ref{table:energy_mean}. 

We see from this data that, as expected, the viscous effects result in energy loss in all cases discussed. The sign of the total energy extracted therefore depends on the relative magnitudes of the energy extracted from the vorticity-induced force and the energy lost to viscous dissipation. For the total energy extraction to be positive, $E^*_\omega$ needs to be large enough to overcome the losses to viscosity. This confirms that the primary driving force for energy extraction is the vorticity in the fluid. Further, we see that the magnitude of $E^*_\omega$ depends strongly on oscillation amplitude as well as the aspect-ratio. In the case of $AR=1.15$ and $A^*_y=0.05$, this positive energy extraction from vorticity is responsible for the overall positive energy extraction. However, there is a significant drop in $E^*_\omega$ on increasing the aspect-ratio slightly to $AR=1.2$ at the same amplitude of oscillation. This drop in the energy extracted from the vorticity-induced forces hence allows the dissipative nature of viscous effects to push the total energy extraction below zero. On increasing the oscillation amplitude to $A^*_y=0.10$, a large increase in $E^*_\omega$ is observed for a cylinder with the same aspect ratio, $AR=1.2$. In fact, $E^*_\omega$ increases by a little more than an order of magnitude on doubling the amplitude of oscillation from $A^*_y=0.05$ to $A^*_y=0.10$ while keeping the aspect-ratio constant at $AR=1.2$. The energy loss to viscous effects however does not show this disproportionate increase, and $E^*_\omega$ hence pushes the total energy extraction to a positive value.

\subsection{Shear-layer and wake contributions to sustained vibration}
\begin{figure}
  \centerline{\includegraphics[scale=1.0]{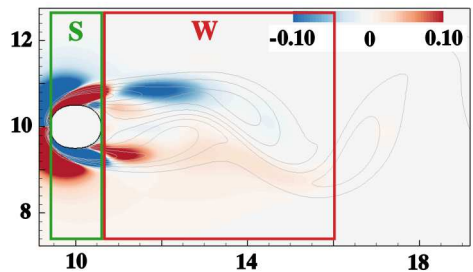}}
  \caption{ Schematic of the integration volumes used to calculate the vorticity-induced force contributions of the shear-layer (denoted by $S$) and wake (denoted by $W$). These integration volumes are overlaid on an instantaneous snapshot of contours of $\vif$ for a case with $AR=1.15$ and $A^*_y=0.05$. }
\label{fig:vif_vol}
\end{figure}
A useful question to ask at this juncture is -- what physical mechanisms and flow phenomena contribute to this complicated dependence of $E^*_\omega$ on the cylinder aspect-ratio and oscillation amplitude? As pointed out before, the volume-integral form of the vorticity-induced force in equation \ref{eq:vif} allows us to isolate the contribution of vorticity in different spatial regions in the flow, to the total force on the cylinder. This allows us to decompose $E^*_\omega$ into contributions from the work done by the vorticity-induced force from these different spatial regions, on the oscillating cylinder. Here, we specifically focus on the effects of the layer of vorticity over the surface of the cylinder (we refer to this as the ``shear layer'') and the vortex wake behind the cylinder. We quantify these effects by choosing domains of integration that isolate the vorticity on the surface of the cylinder from that in the wake of the cylinder. In \ref{fig:vif_vol} we show a schematic of these integration volumes, overlaid on a snapshot of contours of $\vif$ for a case with $AR=1.15$. The rectangular regions denoted by $S$ and $W$ are the integration volumes used in this work for the contributions from the shear layer and wake, respectively. We should note that the integration volumes used here are simply meant to isolate these two mechanisms with the least ambiguity, owing to the lack of strict definitions of the ``shear-layer'' or ``wake'' regions. Hence, since we are interested in isolating the layer of vorticity over the surface of the cylinder, the streamwise extent of the ``shear-layer'' region is defined to span the horizontal length of each cylinder's surface. The region downstream of this is termed the ``wake''. The force contributions from these volumes (which we denote as $C_S$ and $C_W$ respectively) can be computed using equation \ref{eq:vif}, and as before, we compute the energy extracted by the cylinder from the flow in these volumes by using equation \ref{eq:energy_trans}.

\begin{figure}
  \centerline{\includegraphics[scale=1.0]{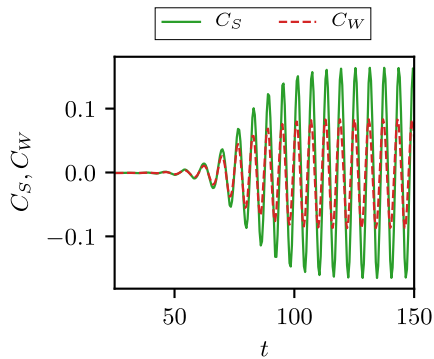}}
  \caption{ Partitioning of vorticity-induced lift force into contributions from the shear layer ($C_S$; green, solid line) and wake ($C_W$; red, dashed line) for stationary cylinders with $AR=1.15$.}
\label{fig:shrl_wake_static}
\end{figure}
A sample partitioning of the vorticity-induced force contributions to lift from the wake and shear-layer is shown for a \textit{stationary} $AR=1.15$ cylinder in figure \ref{fig:shrl_wake_static}. It is interesting to note that the amplitude of force due to the shear-layer is significantly higher than that due to the wake in this case. In fact, this is seen to be true for all the aspect-ratios (including $AR=1.0)$ studied in this work. Phenomenologically, this is due to the fact that the wake vortices have lower vorticity magnitudes than the vorticity layer on the surface and they are also farther away from the transverse surfaces of the cylinder than the vorticity layer on the body. This observation will be relevant to our subsequent discussion about the mechanisms driving the onset and growth of flow-induced vibrations.

\begin{figure}
  \centerline{\includegraphics[scale=1.0]{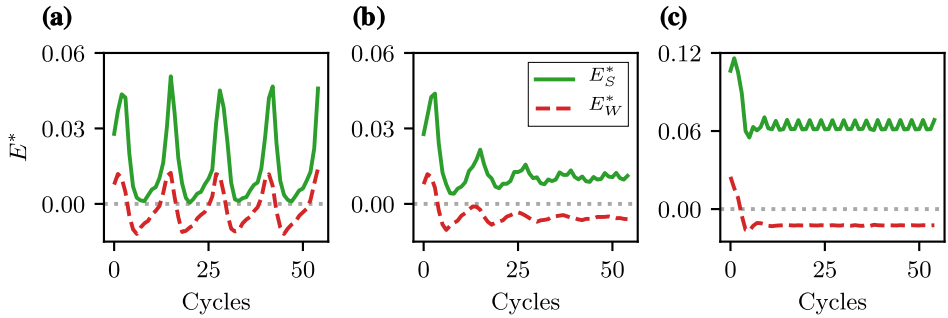}}
  \caption{ Contributions of energy extraction from vorticity-induced force in the shear layer (green, solid line) and wake (red, dashed line), for cylinders of different aspect ratios and oscillating with different amplitudes; (a) $AR = 1.15$, $A^*_y=0.05$; (b) $AR = 1.20$, $A^*_y=0.05$; (c) $AR = 1.20$, $A^*_y=0.10$.}
\label{fig:energy_vif}
\end{figure}
We can now use the above method to examine the mechanisms that cause the complex dependence of $E^*_\omega$ on the aspect-ratio and oscillation amplitude, shown in figure \ref{fig:energy_fpm}. We compare the energy extracted from the vorticity in the shear layer and wake in figure \ref{fig:energy_vif}, for the stationary states of the three cases discussed above. The energy extraction from the shear layer is denoted by $E^*_S$ and that from the wake is denoted by $E^*_W$. For the case with $AR=1.15$ and $A^*_y=0.05$, we see that both $E^*_S$ and $E^*_W$ exhibit low-frequency oscillations due to the non locked-in state of the system, as was seen before. The mean energy transfer from the shear layer and wake in this case are $E^*_S=1.55 \times 10^{-2}$, and $E^*_W=-1.78 \times 10^{-3}$. Hence, we see that the shear layer contributes energy to the oscillation, while the wake vortices extract energy from the oscillation. 

For the case with $AR=1.2$ and $A^*_y=0.05$ too, we find that the shear layer contributes positive energy at the stationary state, with $E^*_S=1.05 \times 10^{-2}$, and the wake vortices extract energy, with $E^*_W=-5.67 \times 10^{-3}$. Note that the shear layer energy extraction in this case ($AR=1.2$) is slightly lower than that in the case with $AR=1.15$, however the energy loss in the wake is approximately $3\times$ higher. The case with $AR=1.2$ and larger amplitude of $A^*_y=0.10$ also shows energy loss due to the wake, along with significantly higher energy extraction in the shear layer. For this case $E^*_S=6.37 \times 10^{-2}$, which represents a roughly $6\times$ increase from the case with the same aspect-ratio and smaller amplitude of oscillation. The energy lost to the wake is $E^*_W=-1.25 \times 10^{-2}$, which is about $2\times$ of the value computed in the case with smaller oscillation amplitude. A summary of these average energy extraction values is provided in table \ref{table:energy_mean}.

An important observation from this spatial partitioning of the energy extraction due to the vorticity-induced force is that for all the cases discussed here, the wake vortices actually extract energy \textit{from} the oscillating cylinder during sustained oscillations. It is therefore the positive work done by the shear-layer that is the driving force for sustained oscillations. This is somewhat contrary to conventional notions in the field of flow-induced vibrations, where significant interest has been directed at the vortex shedding patterns in the wake and their relation to the flow-induced vibration response. 

This spatial partitioning also allows us to identify the reason behind the change in the sign of $E^*$ for a small increase in the cylinder aspect-ratio (only about a $4\%$ increase) - and the corresponding bifurcation in the amplitude response discussed in section \ref{sec:flow_induced} 
This change in $E^*$ is caused by the fact that the energy extracted from the shear layer reduces slightly on increasing $AR$, while the energy lost to the wake increases dramatically. The small reduction in $E^*_S$ is associated with slightly weaker shear layers over the longer-$AR$ cylinder, which is expected. The approximately $3\times$ increase in energy loss due to the wake can be related to a change in the phase difference between the wake and the oscillation as $AR$ is increased. Furthermore, this partitioning also suggests why the region of $E^*<0$ is restricted to low oscillation amplitudes. We see that doubling the oscillation amplitude at this $AR=1.20$ also approximately doubles $E^*_W$. However, the energy transfer due to the shear layer increases dramatically, owing to stronger shear layers as a result of larger oscillation amplitude. In summary, this analysis suggests that the low-amplitude stable branch of the equilibrium curve is caused by the rapid detuning of the wake as the aspect-ratio of the cylinder is increased at small amplitude. Further, the low-amplitude nature of this branch is related to the fact that the energy extracted from the shear layer varies non-linearly with oscillation amplitude, and offsets the dissipative effect of the wake at sufficiently large amplitudes. 

\subsection{Role of the shear-layer and wake in the initiation of oscillations}



The above observation that the wake takes energy away from the oscillation and the shear layer does positive work to sustain stationary-state oscillations brings up the question regarding the \textit{onset} of flow-induced oscillations. The onset cannot be explained by the same mechanism driving sustained oscillations, i.e. energy extraction driven exclusively by the shear layer. This is because the symmetry-breaking in the wake is known to be responsible for the oscillations in the shear layers, and the resultant oscillating lift-force \citep{Zdravkovich1981ReviewShedding,Triantafyllou1986OnCylinders}. Hence we expect that the wake plays a role in facilitating the onset of oscillations on account of this symmetry-breaking. Moreover, since we are using forced oscillations as a proxy for the stationary state of flow-induced oscillations, it is essential to verify that the above findings are not an artefact of forced oscillation and are indeed relevant in sustaining flow-induced oscillations. To  explore the mechanisms that govern the dynamics at their onset and also confirm that the previous observations are not specific to forced oscillations, we now analyze the energy partitioning into the wake and shear-layer contributions for two representative flow-induced oscillation cases.


\begin{figure}
  \centerline{\includegraphics[scale=1.0]{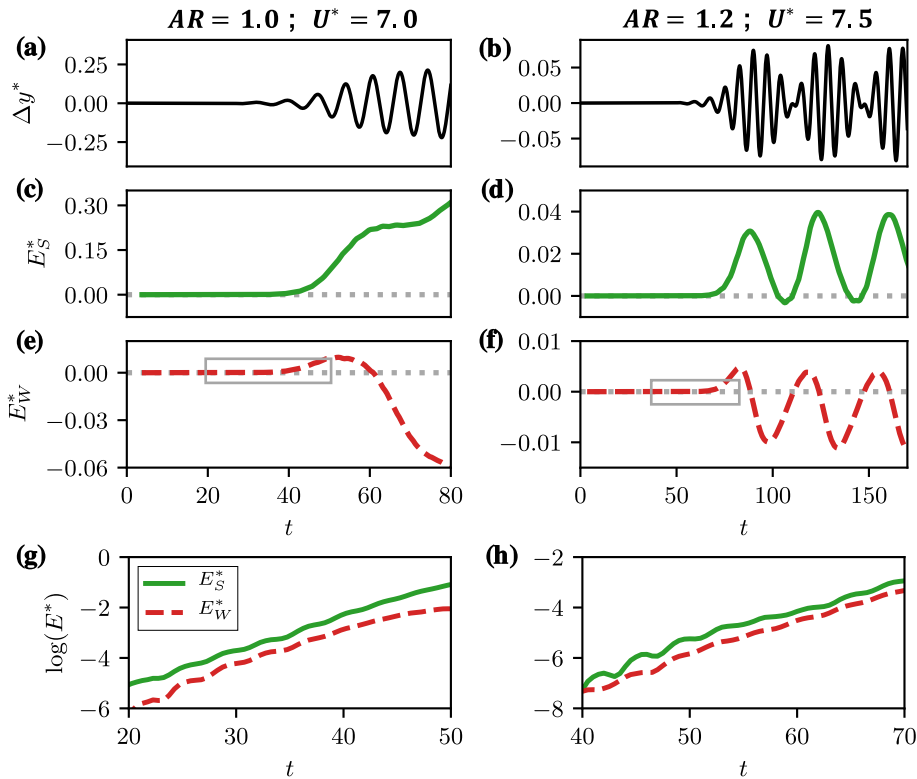}}
   \caption{ Analysis of energy extraction from the shear-layer and wake for two flow-induced oscillation cases very close to the onset of oscillations. Left panel shows a case with $AR = 1.0$, $U^*=7.0$ and right panel shows a case with $AR = 1.20$, $U^*=7.5$; (a-b) Amplitude of heave oscillations ($\Delta y^*$); (c-d) Energy extraction from $\vif$ in the shear-layer ($E^*_S$); (e-f) Energy extraction from $\vif$ in the wake ($E^*_W$); (g-h) Zoom-in of energy contributions from $E^*_S$ and $E^*_W$, plotted on a semi-log scale for the time-period indicated by the grey boxes in figures (e) and (f).}
\label{fig:energy_vif_fim}
\end{figure}
In figure \ref{fig:energy_vif_fim} we plot the energy extracted from the wake ($E^*_W$) and shear layer ($E^*_S$) by two flow-induced oscillating cylinders with different aspect-ratios. The $U^*$ values for these cases ($U^*=7.0$ and $U^*=7.5$) correspond to those around the bifurcation analyzed in \ref{fig:heave_re100}(b). We are particularly interested in the energy balance very close to the onset, and initial growth, of the oscillation. The left-panel of figure \ref{fig:energy_vif_fim} shows this early-time heave amplitude and energy partitioning for a case with $AR=1.0$ and $U^*=7.0$, which shows locked-in oscillations. The time-series plot of heave amplitude in figure \ref{fig:energy_vif_fim}(a) shows that the onset of oscillation occurs at $t \approx 40$. We see that the energy extraction from both the shear-layer and wake, shown in figures \ref{fig:energy_vif_fim}(c) and \ref{fig:energy_vif_fim}(e), are initially zero before the onset of oscillation. However at the onset, the cylinder initially extracts positive energy from both the shear-layer and the wake. As the oscillations grow, the cylinder continues to extract energy from the shear layer at an increasing rate but the wake now becomes an energy-sink for the oscillations. These observations provide confirmation for our earlier forced-oscillation findings from figure \ref{fig:energy_vif}, relating to the competing effects of the shear-layer and wake at finite oscillation amplitudes. Further it also shows that the vortex shedding in the wake does in fact, play a positive role at the onset of oscillation, which is a manifestation of symmetry-breaking.

In the right-panel of figure \ref{fig:energy_vif_fim} we provide further evidence to support these observations. In this case, with $U^*=7.5$ and $AR=1.20$, the oscillations are not in a locked-in state. This leads to a ``beating'' phenomenon in the oscillation as shown in the time-series of heave amplitude in figure \ref{fig:energy_vif_fim}(b). The beats are evident in the energy extraction as well, seen as oscillations in both $E^*_S$ and $E^*_W$, in figures \ref{fig:energy_vif_fim}(d) and \ref{fig:energy_vif_fim}(f) respectively. The beating phenomenon is particularly interesting for this discussion as it represents multiple ``onsets'' of oscillation. We see that each time the oscillation amplitude decreases close to zero, there is a brief period of (positive) extraction of energy from the wake. Subsequent increase in oscillation amplitude leads to a ``detuning'' of the wake, resulting in the wake becoming an energy sink. Further, each instance of amplitude growth is also accompanied by positive energy extraction from the shear layer. However in this case, the absence of frequency lock-in causes a drift in the phase difference between the oscillation and vortex shedding. This unfavourable phase difference in the wake is able to feedback into the shear layer at these small oscillation amplitudes ($E^*_S$ and $E^*_W$ have comparable magnitudes), resulting in an increased detuning of the shear layers, thereby reducing the energy extraction from the shear layers. The periodicity of these mechanisms causes this process to repeat and hence leads to beats in the oscillation amplitude. This is an interesting demonstration of the competing influence of the wake and shear-layer in producing variations in the oscillation amplitude. 


From the above cases we note that while at large amplitudes the shear-layer is the driving force for the oscillations, at smaller amplitude, the wake vortices continue to play an important role. Close to the onset of oscillations in particular, both these mechanisms do positive work on the cylinder. To provide a clearer picture of the relative importance of these two mechanism at the onset of flow-induced oscillations, we examine the contributions of wake vortices and shear layers very early in the initial amplitude growth for the two cases discussed above. In this analysis, we focus on the time-period indicated by the grey boxes in figures \ref{fig:energy_vif_fim}(e) and \ref{fig:energy_vif_fim}(f). The energy contributions from the shear-layer and wake, plotted on a semi-log scale for the two flow-induced oscillation cases discussed above, are shown in figures \ref{fig:energy_vif_fim}(g) and \ref{fig:energy_vif_fim}(h). For both cases, $U^*=7.0$ and $AR=1.0$ in figure \ref{fig:energy_vif_fim}(g) as well as $U^*=7.5$ and $AR=1.2$ in figure \ref{fig:energy_vif_fim}(h), we see that the shear-layer is the larger contributor of positive energy even at this very early time. This is in line with our earlier observation for a static cylinder (see figure \ref{fig:shrl_wake_static}), where we saw that the shear-layer generates a larger contribution to the lift force than the wake vorticity. Using this static condition as a proxy for the pre-onset state of the oscillator, figures \ref{fig:shrl_wake_static} and \ref{fig:energy_vif_fim}(g-h) are indicative of the same phenomenon - that it is the pressure variation generated directly by the shear-layers that is the dominant driving mechanism even very close to the onset of oscillation. However we should reiterate that, as mentioned above, the symmetry of the problem requires vortex shedding in the wake to initiate oscillations in the shear-layer forcing. Hence although the shear-layer is the \textit{dominant} mechanism very close to the onset of oscillations, it cannot act independently to initiate the oscillations.

\subsection{Underlying mechanism for the shear-layer contribution to oscillations}
\newcommand\numberthis{\addtocounter{equation}{1}\tag{\theequation}}

\begin{figure}
  \centerline{\includegraphics[scale=1.0]{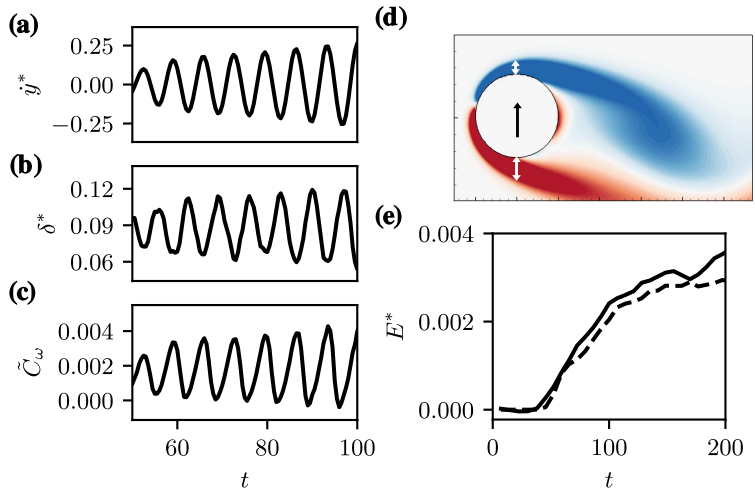}}
  \caption{ Comparison of heave velocity, boundary-layer thickness, and shear-layer forcing for a case of flow-induced oscillations with $AR=1.0$ and $U^*=7.0$ (a) Time-series of heave velocity ($\dot{y}^*$); (b) Boundary-layer displacement thickness ($\delta^*$) on the top surface of the cylinder, along the transverse axis; (c) Force contribution from the dominant terms of $\vif$ in the shear-layer on the top surface of the cylinder ($\tilde{C}_{\omega}$; equation \ref{eq:vif_terms}); (d) Instantaneous snapshot of vorticity contours around the cylinder as it moves upward, with white arrows showing qualitative measure of boundary-layer thickness on the two transverse surfaces; (e) Comparison of total energy extracted from $\vif$ (solid line) with that extracted from $\tilde{C}_{\omega}$ (dashed line) in a small region inside the top-surface boundary-layer.}
\label{fig:bl_thickness}
\end{figure}

As a final piece in this analysis, we seek to provide a physical basis for the net-positive energy contribution of the shear-layer for all the cases simulated here. We hypothesize that the pressure force induced by the shear layers is directly related to changes in the thickness of the shear layers on the transverse surfaces of the cylinder, which in turn, is driven by the transverse oscillations of the cylinder. Such a relationship would explain two key findings in the above analysis - that the shear-layer forcing is at a favourable phase with respect to the heave velocity of the cylinder, and also that its influence strongly depends on the oscillation amplitude. In order to assess this hypothesis, we compare the boundary layer thickness on the top surface of the oscillating cylinder to the transverse velocity, and also relate the changes in the boundary-layer thickness to the fluctuations in $\vif$ associated with the shear-layer. This comparison is performed for the case of flow-induced oscillations with $AR=1.0$ and $U^*=7.0$, which was discussed above. 

In figure \ref{fig:bl_thickness}(a) we show the time-series of heave-velocity for this case, during a time-period when the amplitude of oscillation is growing. To compare the heave-velocity with the boundary-layer thickness, we compute the displacement thickness on the top-surface of the cylinder along the transverse axis. The displacement thickness is calculated as 
\begin{equation}
    \delta^* = \int_{y_{min}}^{y_{max}}\Big( 1- \frac{u}{U_{max}} \Big) dy
\end{equation}
where $U_{max}$ is the maximum horizontal velocity in the shear-layer, $y_{min}$ is the $y-$coordinate of the top surface of the cylinder, $y_{max}$ is the $y-$coordinate at which $U_{max}$ occurs. The computed displacement thickness is plotted against time in figure \ref{fig:bl_thickness}(b), and we see that the displacement thickness follows an inverse relationship to the heave-velocity. This confirms that the boundary layer on the transverse surface of the cylinder compresses(expands) with the upward(downward) motion of the cylinder. As further qualitative evidence of this effect, figure \ref{fig:bl_thickness}(d) shows an instantaneous snapshot of vorticity contours around the cylinder as it moves upward, and the white arrows show that the boundary layer on the upper surface is thinner than that on the lower surface. 

In order to relate this boundary-layer thickness to the vorticity-induced force contained in the shear-layer, we note that the mathematical form for $\vif$ given in equation \ref{eq:vif} can be expanded into numerous terms involving the gradients of the horizontal and vertical velocity components in the flow. We estimate the importance of each of these terms in driving the oscillations by probing the flow in a small region within the shear-layer and computing the energy extracted by the cylinder from each term. The integration volume used for this exercise is taken as one grid cell, located along the axis of the heave motion at $0.1D$ above the surface of the cylinder (moving with the cylinder). We find that two terms in particular, shown below, correspond to the bulk of the energy extraction in this region:
\begin{align*}
    \vif \approx \tilde{C}_{\omega} &= \int_\vol \bigg[ \Big( u_v \frac{\partial u_v}{\partial y} + u_v \frac{\partial u_{\Phi}}{\partial y} + u_{\Phi} \frac{\partial u_v}{\partial y} \Big)\frac{\partial \phi}{\partial y} \bigg] dV \\ 
    &+ \int_\vol \bigg[ \frac{\partial}{\partial y} \Big( v_v \frac{\partial v_v}{\partial y} + v_v \frac{\partial v_{\Phi}}{\partial y} + v_{\Phi} \frac{\partial v_v}{\partial y} \Big)\phi \bigg] dV. 
    \label{eq:vif_terms} \numberthis
\end{align*}
Here $u_v$ and $u_{\Phi}$ are the rotational and curl-free components of the horizontal velocity, and $v_v$ and $v_{\Phi}$ are the corresponding components of the vertical velocity. The fact that the approximation $\tilde{C}_{\omega}$ in equation \ref{eq:vif_terms} dominates the energy extraction is confirmed in figure \ref{fig:bl_thickness}(e) by comparing the total energy extraction from $\vif$ in this small region (solid line) with the energy accounted for by $\tilde{C}_{\omega}$ (dashed line). It is noteworthy that these terms are primarily composed of vertical derivatives of the velocity components. This suggests that the vorticity-induced force in the shear-layer is indeed driven by the fluctuations in the boundary-layer thickness. This can be verified by comparing fluctuations in $\tilde{C}_{\omega}$, which is plotted in figure \ref{fig:bl_thickness}(c) with $\delta^*$. We see that minima in $\delta^*$ correspond to maxima in $\tilde{C}_{\omega}$ as expected. The fact that $\delta^*$ is in-turn inversely related to the transverse velocity ($\dot{y}^*$) then explains why the shear-layer has a net-positive influence on the oscillation for all the cases discussed here. This supports our hypothesis regarding the underlying reason for the dominant influence of the shear-layer in sustaining flow-induced vibrations.

\subsection{Phenomenology of flow-induced oscillations}
In summary, the above analysis of energy extraction in flow-induced as well as forced oscillations suggests the following phenomenology for the initiation and sustenance of the flow-induced vibration of cylinders:
\begin{enumerate}
  \item Symmetry breaking in the wake results in vortex shedding, which also induces periodic oscillation in the attached boundary/shear layers.
  \item The oscillation in the boundary/shear layer as well as the periodically shed wake vortices induce oscillations in the lift force on the cylinder. Both these effects play a role in  initiating the cylinder oscillations. However, the direct contribution of the boundary/shear layer to the oscillating lift and amplitude growth is larger than the contribution of the wake vortices. 
  \item For oscillations at small amplitudes, the shear-layer is affected both by the oscillation of the cylinder as well as the wake vortex shedding, and the frequency of these two might be different. In this case, the pressure force induced by the shear layer might detune from the cylinder oscillation periodically, resulting in a beating phenomenon in the cylinder motion.
  \item Once the oscillations grow beyond a certain amplitude, the lift contribution of the wake vortices goes out of phase with the cylinder oscillation, and wake vortices begin to act as sinks of energy. In contrast, the shear layer is driven primarily by the cylinder motion and the contribution to energy extraction from the flow increases with oscillation amplitude. It is this mechanism that sustains stationary state oscillations against the constant drain of energy by the vortex shedding in the wake. Further, this increase in shear-layer forcing with amplitude is governed by thickening and thinning of the boundary-layers on the transverse surfaces of the cylinder, which is a consequence of the heaving motion. 
  
\end{enumerate}

\section{Conclusions}
\label{sec:conclusions}

In this work, we use a force partitioning method in conjunction with an energy-based analysis to dissect the contributions of different fluid forcing mechanisms - such as viscous effects, added-mass force, and vorticity-induced force - on the flow-induced vibration of cylinders. The work done on the oscillating cylinder by each of these forcing mechanisms, or the energy extracted by the cylinder, provides rigorous quantification of their effect on the oscillation. Further, the force partitioning method also allows us to isolate the contributions of different spatial regions of the flow towards the total vorticity-induced force. These methods therefore allow us to quantify the contributions of specific flow mechanisms/features in driving the flow-induced vibration of cylinders. The specific mechanisms analyzed in this work are the effect of the shear-layer and the vortex-wake in initiating and sustaining oscillations. 

We show quantitatively that the driving factor behind the flow-induced oscillations is the vorticity-induced force. For the cases analyzed here, viscous effects are always seen to dissipate energy from the oscillation. Thus, the work done by the vorticity-induced component of the total force needs to overcome these viscous effects in order to generate sustained oscillations. By further decomposing the energy extracted from vorticity into the specific contributions of the shear layer and the wake, we show that while the oscillations are briefly energized at their onset by the vortex shedding in the wake, the growth as well as sustenance of these oscillations is driven directly not by the wake vortices but by the shear layer on the transverse surfaces of the cylinder. In fact, beyond the onset phase, wake vortices actually act as sinks of energy during sustained oscillation. We highlight these findings in the context of the effect of aspect-ratio on flow-induced oscillations of a cylinder and explain the rapid drop in oscillation amplitude with a minor increase in aspect ratio.

Indeed, the conventional use of the term ``vortex-induced vibration" to describe the overall phenomenon has an inherent implication that the oscillations of elastically mounted bluff bodies are induced by the wake \textit{vortices} but our analysis shows that this is not the case. Indeed, ``vorticity-induced vibration" is a more apt descriptor since first, vorticity-induced lift forces dominate over all other possible contributions, and second, it is the \textit{vorticity} in the shear layers and not the \textit{vortices} in the wake that sustain the oscillations. 


The application of the force partitioning method in combination with the energy analysis demonstrates the utility of these methods in the study of fluid-structure interaction problems. In addition, the force partitioning method used here is also relevant in evaluating the forces on immersed bodies in applications that lack direct access to the pressure field - such as in some experiments, and also computational methods formulated in non-primitive variables. We should note that the accuracy of this force partitioning method relies on well-resolved flow-field data. However, modifications of this method \citep{Pan2002AFlow}, as well as the development of related methods to evaluate forces based on flow-field integrals \citep{Wu1981TheoryFlows,Noca1997MeasuringDerivatives,Noca1999ADerivatives} have made these methods more amenable to under-resolved flow-field data.

This analysis also brings up several important questions. We show in this work that the energy extracted from the vortex wake and shear layer change dramatically with small changes in cylinder aspect-ratio and oscillation amplitude and this could form the basis of shape modification for vibration control. Another potential extension of this work could involve the effect of Reynolds number and three-dimensionality on the observed phenomena. Further, while this work has exclusively focused on transverse vibrations due to the fact that they are generally observed to occur with larger amplitudes than in-line oscillations \citep{Singh2005Vortex-inducedModes,Prasanth2008Vortex-inducedNumbers,Bearman2011CircularVibrations,Navrose2014FreeNumbers}, the latter does have an affect on the dynamics of transverse oscillations. Hence the study of the forcing mechanisms and flow-induced response of the system studied here, with the inclusion of in-line oscillations, is a possible extension of this work.

\section*{Acknowledgments}
This work is supported by the Air Force Office of Scientific Research Grant Number FA 9550-16-1-0404, monitored by Dr. Gregg Abate. This work also benefited from the computational resources at Extreme Science and Engineering Discovery Environment (XSEDE), which is supported by National Science Foundation grant number ACI-1548562, through allocation number TG-CTS100002. Computational resources at the Maryland Advanced Research Computing Center (MARCC) are also acknowledged.

\section*{Declaration of interests}
The authors report no conflict of interest.

\appendix

\section{Force partitioning method}
\label{app:fpm_derivation}

\renewcommand{\kin}{C^{(i)}_\kappa}
\renewcommand{\vif}{C^{(i)}_\omega}
\renewcommand{\shr}{C^{(i)}_\sigma}
\renewcommand{\pot}{C^{(i)}_\Phi}
\renewcommand{\ext}{C^{(i)}_\Sigma}

In this section, we provide additional details about the derivation of the terms in the force partitioning shown in equations \ref{eq:kin} to \ref{eq:ext}. We note that some aspects of the discussion in section \ref{sec:fpm} are repeated here for the sake of a self-contained derivation.

We first define some nomenclature and the general setup relevant to the force partitioning. We are  interested in partitioning the total force on a body in the $i$-direction. The surface of the body is represented  by $B$, and it is immersed in a fluid domain of volume $\vol$. The spatial domain is bounded externally by the surface $\Sigma$, and the unit vector $\vec{n}$, defining the orientation at every point along the bounding surfaces $B$ and $\Sigma$, points out from the fluid volume.

We start the derivation with the Navier-Stokes momentum equation, written in Lamb-Gromeka form,
\begin{equation}
  \frac{\partial \vec{u}}{\partial t} + \vec{\omega} \times \vec{u} + \frac{1}{2}\vec{\nabla}(\vec{u} \cdot \vec{u})= - \vec{\nabla} p - \frac{1}{Re} \vec{\nabla} \times \vec{\omega} 
  \label{eq:lamb-gromeka}
\end{equation}

As discussed in section \ref{sec:fpm}, we construct an auxiliary potential, $\phi^{(i)}$,  at every time-instance, which is a function of the instantaneous position and shape of the immersed body as well as the outer domain boundary. This potential is defined as:
\begin{equation}
  \vec{\nabla}^2 \phi^{(i)} = 0, \ \ \mathrm{ with } \ \
  \vec{n} \cdot \vec{\nabla} \phi^{(i)}=
    \begin{cases}
      n_i \;, \; \mathrm{on} \; B \\
      0 \; \;, \; \mathrm{on} \; \Sigma \\
    \end{cases}
  \label{eq:scalar_deriv}
\end{equation}
It is clear from equation \ref{eq:scalar_deriv} that $\vec{\nabla} \phi^{(i)}(\tau)$ is the instantaneous potential flow-field (or ideal flow) around a body $B$ that is moving with unit velocity in the $i$-direction. 




The Navier-Stokes equation (\ref{eq:lamb-gromeka}) is now projected on to the gradient of this auxiliary potential, and the result is integrated over the volume of the fluid domain, $\vol$:
\begin{align*}
  \int_\vol \frac{\partial \vec{u}}{\partial t} \cdot \vec{\nabla} \phi^{(i)} dV &+ \int_\vol (\vec{\omega} \times \vec{u}) \cdot \vec{\nabla} \phi^{(i)} dV + \int_\vol \frac{1}{2}\vec{\nabla}(\vec{u} \cdot \vec{u}) \cdot \vec{\nabla} \phi^{(i)} dV \\
  &= -\int_\vol \vec{\nabla} p \cdot \vec{\nabla} \phi^{(i)} dV - \int_\vol \frac{1}{Re} (\vec{\nabla} \times \vec{\omega}) \cdot \vec{\nabla} \phi^{(i)} dV \numberthis
  \label{eq:projection}
\end{align*}
We will now simplify each term of the above equation separately, starting with the pressure term on the right-hand side.
\begin{equation}
   \int_\vol \vec{\nabla} p \cdot \vec{\nabla} \phi^{(i)} dV  = \int_\vol \vec{\nabla} \cdot (p \vec{\nabla} \phi^{(i)}) dV 
  \label{eq:pressure_1}
\end{equation}
Using the divergence theorem, 
\begin{equation}
   \int_\vol \vec{\nabla} \cdot (p \vec{\nabla} \phi^{(i)}) dV = \int_{B+\Sigma} p \vec{n} \cdot \vec{\nabla} \phi^{(i)} dS = \int_{B} p n_i dS
  \label{eq:pressure_final}
\end{equation}
where the last step follows from the boundary condition on the field $\phi^{(i)}$. For $\vec{n}$ pointing from the fluid into the surface of the immersed body, this is evidently the pressure force on the body exerted by the surrounding fluid.
We simplify the third term on the left-hand side in a similar manner. Hence,
\begin{equation}
  \int_\vol \frac{1}{2}\vec{\nabla}(\vec{u} \cdot \vec{u}) \cdot \vec{\nabla} \phi^{(i)} dV = \int_B \frac{1}{2}(\vec{u} \cdot \vec{u}) n_i dS
  \label{eq:u_dot_u}
\end{equation}
We now treat the viscous term as follows:
\begin{align*}
  \int_\vol \frac{1}{Re} (\vec{\nabla} \times \vec{\omega}) \cdot \vec{\nabla} \phi^{(i)} dV = \int_\vol \frac{1}{Re} \vec{\nabla} \cdot ( \vec{\omega} \times \vec{\nabla} \phi^{(i)}) dV \\ 
  = \int_{B+\Sigma} \frac{1}{Re} \vec{n} \cdot ( \vec{\omega} \times \vec{\nabla} \phi^{(i)}) dS = \int_{B+\Sigma} \frac{1}{Re} (\vec{n} \times \vec{\omega}) \cdot \vec{\nabla} \phi^{(i)} dS \numberthis
  \label{eq:viscous}
\end{align*}
Finally, we simplify the unsteady term. Using incompressibility ($\vec{\nabla} \cdot \vec{u}=0$) and the divergence theorem,
\begin{equation}
  \int_\vol \frac{\partial \vec{u}}{\partial t} \cdot \vec{\nabla} \phi^{(i)} dV = \int_\vol \vec{\nabla} \cdot \bigg( \frac{\partial \vec{u}}{\partial t} \phi^{(i)} \bigg) dV = \int_{B+\Sigma} \vec{n} \cdot \bigg( \frac{\partial \vec{u}}{\partial t} \phi^{(i)} \bigg) dS
  \label{eq:unsteady_1}
\end{equation}
Further, using the definition of the total derivative $d/dt$,
\begin{align*}
  \int_{B+\Sigma} \vec{n} \cdot \bigg( \frac{\partial \vec{u}}{\partial t} \phi^{(i)} \bigg) dS = \int_{B+\Sigma} \vec{n} \cdot \Bigg [ \bigg\{ \frac{d \vec{u}}{dt} - (\vec{\omega} \times \vec{u}) - \frac{1}{2}\vec{\nabla}(\vec{u} \cdot \vec{u}) \bigg\}\phi^{(i)} \Bigg] dS \\
  = \int_{B+\Sigma} \vec{n} \cdot \bigg( \frac{d \vec{u}}{dt} \phi^{(i)} \bigg) dS - \int_\vol \vec{\nabla} \cdot \Bigg [ \bigg\{ (\vec{\omega} \times \vec{u}) + \frac{1}{2}\vec{\nabla}(\vec{u} \cdot \vec{u}) \bigg\}\phi^{(i)} \Bigg] dV \numberthis
  \label{eq:unsteady_final}
\end{align*}

We now plug equations \ref{eq:pressure_final}, \ref{eq:u_dot_u}, \ref{eq:viscous} and \ref{eq:unsteady_final} into equation \ref{eq:projection}. Performing some further simplification and rearranging terms,
\begin{align*}
  \int_{B} p n_i dS = &- \int_{B} \vec{n} \cdot \bigg( \frac{d \vec{u}}{dt} \phi^{(i)} \bigg) dS - \int_B \frac{1}{2}(\vec{u} \cdot \vec{u}) n_i dS + \int_\vol \frac{1}{2} \vec{\nabla} \cdot \Bigg [ \bigg\{ \vec{\nabla}(\vec{u} \cdot \vec{u}) \bigg\}\phi^{(i)} \Bigg] dV \\ 
  &+ \int_\vol \bigg[ \vec{\nabla} \cdot (\vec{\omega} \times \vec{u})\bigg] \phi^{(i)} dV  + \int_{B} \frac{1}{Re} (\vec{\omega} \times \vec{n}) \cdot \vec{\nabla} \phi^{(i)} dS \\ 
  &- \int_{\Sigma} \vec{n} \cdot \bigg( \frac{d \vec{u}}{dt} \phi^{(i)} \bigg) dS + \int_{\Sigma} \frac{1}{Re} (\vec{\omega} \times \vec{n}) \cdot \vec{\nabla} \phi^{(i)} dS \numberthis
  \label{eq:pre-final}
\end{align*}

We can gain further insight into the physical relevance of the various terms in the partitioning by separating the contributions of vorticity from the potential flow in the force production. This is done using the Helmholtz decomposition \citep{BATCHELOR} as follows:
\begin{equation}
  \vec{u} = \vec{u}_\Phi + \vec{u}_v = \vec{\nabla} \Phi + \vec{\nabla} \times A
  \label{eq:helmholtz_deriv}
\end{equation}
where $\Phi$ and $A$ are scalar and vector potentials respectively. 

Using this decomposition in equation \ref{eq:pre-final}, and referring to the velocity on the surface of the immersed body as $\vec{U}_B$, i.e. $\vec{u}(B) = \vec{U}_B$, we now arrive at the partitioning of the pressure-force on the body:
\begin{align*}
  \int_{B} p n_i dS &= - \int_{B} \vec{n} \cdot \bigg( \frac{d \vec{U}_B}{dt} \phi^{(i)} \bigg) dS - \int_B \frac{1}{2}(\vec{U}_B \cdot \vec{U}_B) n_i dS   \\
  &+ \int_\vol \Bigg\{ \Big[ \vec{\nabla} \cdot (\vec{\omega} \times \vec{u})\Big] \phi^{(i)} + \vec{\nabla} \cdot \bigg[ \vec{\nabla}\bigg(\frac{1}{2}\vec{u}_v \cdot \vec{u}_v + \vec{u}_v \cdot \vec{u}_{\Phi}\bigg) \phi^{(i)} \bigg] \Bigg\} dV  \\
  &+ \int_{B} \frac{1}{Re} (\vec{\omega} \times \vec{n}) \cdot \vec{\nabla} \phi^{(i)} dS  \\ 
  &+ \int_\vol \vec{\nabla} \cdot \Bigg [ \vec{\nabla}\bigg(\frac{1}{2}\vec{u}_{\Phi} \cdot \vec{u}_{\Phi}\bigg)\phi^{(i)} \Bigg] dV  \\
  &+ \int_{\Sigma} \Bigg\{ -\vec{n} \cdot \bigg( \frac{d \vec{u}}{dt} \phi^{(i)} \bigg) + \frac{1}{Re} (\vec{\omega} \times \vec{n}) \cdot \vec{\nabla} \phi^{(i)} \Bigg\} dS \numberthis
  \label{eq:pressure_decomp_deriv}
\end{align*}

It must be noted that the total force on an immersed body includes shear contributions (which depends on the viscous stress tensor). This shear-force coefficient on the surface $B$ in the $i$-direction can be written in the form $ C'_{\sigma} = -(1/Re) \int_B (\vec{\omega} \times \vec{n}) \cdot \hat{e}_i dS$. Adding this contribution to equation \ref{eq:pressure_decomp_deriv} above, we arrive at the final form for the partitioning of the total force on the body shown in section \ref{sec:fpm}: 
\begin{align}
  C_i &= - \int_{B} \vec{n} \cdot \bigg( \frac{d \vec{U}_B}{dt} \phi^{(i)} \bigg) dS - \int_B \frac{1}{2}(\vec{U}_B \cdot \vec{U}_B) n_i dS  \label{eq:kin_deriv} \\
  &+ \int_\vol \Bigg\{ \Big[ \vec{\nabla} \cdot (\vec{\omega} \times \vec{u})\Big] \phi^{(i)} + \vec{\nabla} \cdot \bigg[ \vec{\nabla}\bigg(\frac{1}{2}\vec{u}_v \cdot \vec{u}_v + \vec{u}_v \cdot \vec{u}_{\Phi}\bigg) \phi^{(i)} \bigg] \Bigg\} dV \label{eq:vif_deriv} \\
  &+ \int_{B} \frac{1}{Re} \bigg[ (\vec{\omega} \times \vec{n}) \cdot \vec{\nabla} \phi^{(i)} - (\vec{\omega} \times \vec{n}) \cdot \hat{e}_i \bigg] dS \label{eq:shr_deriv} \\ 
  &+ \int_\vol \vec{\nabla} \cdot \Bigg [ \vec{\nabla}\bigg(\frac{1}{2}\vec{u}_{\Phi} \cdot \vec{u}_{\Phi}\bigg)\phi^{(i)} \Bigg] dV \label{eq:pot_deriv} \\
  &+ \int_{\Sigma} \Bigg\{ -\vec{n} \cdot  \bigg( \frac{d \vec{u}}{dt} \phi^{(i)} \bigg) + \frac{1}{Re} (\vec{\omega} \times \vec{n}) \cdot \vec{\nabla} \phi^{(i)} \Bigg\} dS \label{eq:ext_deriv}
\end{align}

The terms in equations \ref{eq:kin_deriv} to \ref{eq:ext_deriv} above correspond to the final forms for the force components $\kin$, $\vif$, $\shr$, $\pot$, and $\ext$ respectively, shown in section \ref{sec:fpm}.

Thus the final form of the partitioning can be written compactly as 
\begin{equation}
  C_i = \kin + \vif + \shr + \pot + \ext
  \label{eq:force_decomp_initial}
\end{equation}

\section{Supplementary analysis of aspect-ratio effects}
\label{app:re250}
\label{app:hysteresis}
\label{app:energy_map_15}
\begin{figure}
  \centerline{\includegraphics[scale=1.0]{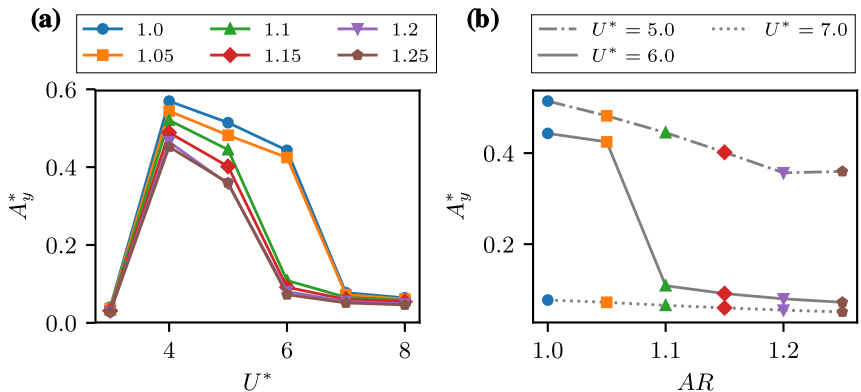}}
  \caption{ Heave amplitude response of flow-induced oscillations at $Re=250$; (a) Maximum heave amplitude versus reduced velocity for different aspect-ratios; (b) Maximum heave amplitude as a function aspect ratio at $U^* = 5.0,6.0,7.0$. }
\label{fig:heave_re250}
\end{figure}
This section briefly describes some supplementary results that are meant to provide further insight into the discussion in section \ref{sec:flow_induced_energy}. We begin with the amplitude response of flow-induced oscillations as a function of $U^*$ and $AR$, for the case of $Re=250$. In figure \ref{fig:heave_re250}(a) we show the maximum heave amplitude, $A^*_y$ as a function of $U^*$, plotted for cylinders of different aspect-ratios using colors and symbols as in figure \ref{fig:heave_re100}. We see the often reported jump in oscillation amplitude within the lock-in regime for all values of $AR$. However the size of this lock-in regime, in terms of $U^*$, changes with aspect-ratio, especially at the high-$U^*$ end of the lock-in regime. We see that this regime extends to $U^*=7.0$ for the cases with $AR=1.0$ and $AR=1.05$, while the lock-in regime ends at $U^*=6.0$ for higher aspect-ratios. This causes a dramatic drop in oscillation amplitude as a function of $AR$ at a fixed reduced velocity of $U^*=6.0$. This is shown in figure \ref{fig:heave_re250}(b) where we compare the heave amplitude for cases with $U^*=5.0$, $U^*=6.0$, and $U^*=7.0$, as a function of $AR$. We see that there is an approximately four-fold drop in oscillation amplitude at $U^*=6.0$, when increasing the aspect-ratio a small amount from $AR=1.05$ to $AR=1.1$.

\begin{figure}
  \centerline{\includegraphics[scale=1.0]{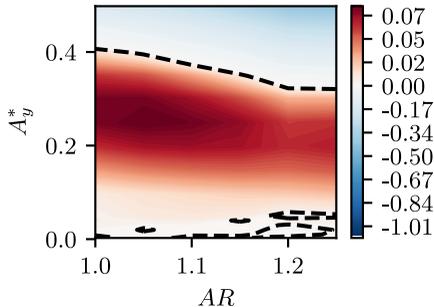}}
  \caption{Contours of energy transfer between the cylinder and flow, as a function of oscillation amplitude ($A^*_y$) and cylinder aspect ratio ($AR$) at $f^*=0.15$ and $Re=100$. This energy transfer is computed using forced oscillations. The dashed line shows the zero energy transfer contour.}
\label{fig:energy_cont}
\end{figure}
As a second aspect of this supplementary analysis, we extend the discussion of the energy map in section \ref{sec:energy_trans}, in order to analyze the robustness of the two low-amplitude equilibrium curves shown in figure \ref{fig:energy_maps}(a). As mentioned in section \ref{sec:energy_trans}, beating oscillations can be written as a sum of sine-waves, and the presence of a beating phenomenon in the flow-induced oscillations can be considered to act like a perturbation to the equilibrium curves predicted by the energy map. This is because for a given aspect-ratio, the energy extraction is computed by performing forced oscillations at a frequency that corresponds to the dominant stationary-state flow-induced oscillation frequency observed at $U^*=7.5$. In the frequency spectrum of flow-induced oscillators that exhibit beats, this dominant frequency is accompanied by a second (smaller) frequency peak. However the energy map shown in figure \ref{fig:energy_maps}(a) neglects the possible energy extracted by the oscillator at this secondary frequency. We find that this second frequency corresponds to approximately $f^* \approx 0.15$ for all the flow-induced oscillation cases at $U^*=7.5$. Hence, in order to examine the effect of this ``frequency perturbation'' on the energy map shown in figure \ref{fig:energy_maps}(a), we compute a second energy map at this fixed frequency of $f^* = 0.15$. We use the same procedure as that described in section \ref{sec:energy_trans}, except that in this case the oscillation frequency is held constant at $f^*=0.15$ for all values of $AR$. Figure \ref{fig:energy_cont} shows the structure of this energy map, where the $E^*=0$ contour line is shown using a dashed line as before. We immediately notice that the equilibrium curve that is present at $1.05 \lessapprox AR \lessapprox 1.10$ in figure \ref{fig:energy_maps}(a) is not present at $f^* = 0.15$ in figure \ref{fig:energy_cont}. However, the equilibrium curve at $AR \gtrapprox 1.20$ persists in both cases. This hence provides additional evidence that the equilibrium curve at $1.05 \lessapprox AR \lessapprox 1.10$, seen in figure \ref{fig:energy_maps}(a), is a weaker and less stable equilibrium than that at $AR \gtrapprox 1.20$.

\section{Solver verification and grid convergence}
\label{app:validation}

\begin{figure}
  \centerline{\includegraphics[scale=0.8]{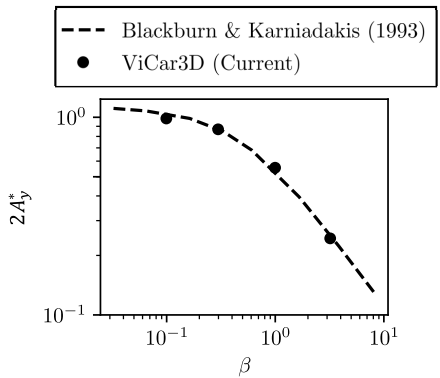}}
  \caption{Flow-induced oscillation amplitude of a circular cylinder versus mass-damping parameter $\beta=8 \pi^2 St^2 m\zeta/\rho D^2$. Results from the present code are shown using symbols and are compared with results from \cite{Blackburn1993Two-Cylinder} shown as a dashed line.}
\label{fig:cyl_validation}
\end{figure}
We have made various comparisons of results from the code used in this work with previous studies, in addition to the validation already mentioned in \cite{Mittal2008ABoundaries}. Specifically, force coefficients as well as Strouhal numbers predicted for flow over static airfoils match very well with existing literature. These results are not shown here for the sake of brevity. In order to validate our fluid-structure interaction solver, we have performed simulations of flow-induced vibration of circular cylinders and compared our amplitude response with the results of \cite{Blackburn1993Two-Cylinder}. In figure \ref{fig:cyl_validation} we show the amplitude response of a freely vibrating cylinder at $Re=200$ with $m^*=m/\rho D^2=10$, against the mass-damping parameter $\beta=8 \pi^2 St^2 m\zeta/\rho D^2$. Here, $St,m,\zeta,\rho$, and $D$ are the vortex shedding Strouhal number, mass, damping ratio, fluid density, and cylinder diameter respectively. The symbols show results from our simulations and the dashed line is from the work of \cite{Blackburn1993Two-Cylinder}. The agreement is quite good.

\begin{figure}
  \centerline{\includegraphics[scale=0.8]{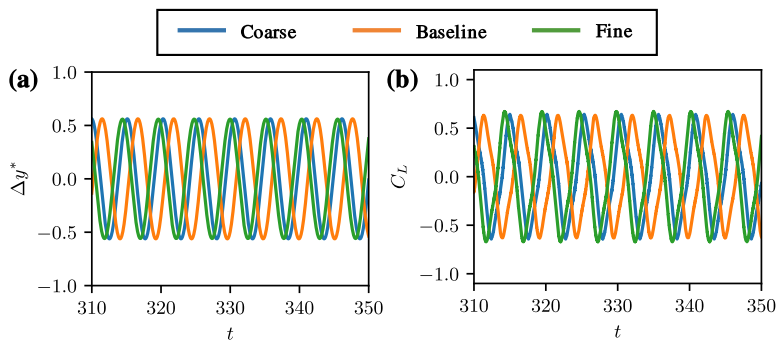}}
  \caption{Comparison of flow-induced oscillation results using three grids to establish grid convergence. Coarse, baseline, and fine grids correspond to $45$, $60$ and $90$ points across the diameter of the cylinder respectively. Here, $Re=100$, $AR=1.0$ and $U^*=5.0$; (a) Timeseries of heave oscillation amplitude; (b) Timeseries of $C_L$ oscillations. }
\label{fig:grid}
\end{figure}
Grid convergence is assessed by performing flow-induced oscillation simulations at $Re=100$, using a circular cylinder ($AR=1.0$) and fixing $U^*=5.0$. The size of the grid used for the results reported in this study is $320 \times 288$ cells, which corresponds to $60$ cells across the diameter of the cylinder. This baseline grid is compared with coarse and fine grids of sizes $288 \times 224$ cells, and $384 \times 384$ cells respectively. These correspond to $45$ and $90$ points across the diameter of the cylinder. In figure \ref{fig:grid}(a) we show the timeseries of heave oscillations computed using the three grids, where the maximum heave amplitude differs by $0.9\%$ between the coarse and baseline grid, and $0.7\%$ between the coarse and fine grid. The timeseries of $C_L$ oscillations are shown in figure \ref{fig:grid}(b), and we see that the r.m.s. differs by $0.6\%$ between the coarse and baseline grids, and $5.3\%$ between the coarse and fine grids. The corresponding differences in the maximum $C_L$ values are $3.8\%$ and $3.0\%$. Hence we consider the computations grid independent. 

\bibliographystyle{jfm}
\bibliography{references}

\begin{thebibliography}{60}
\expandafter\ifx\csname natexlab\endcsname\relax\def\natexlab#1{#1}\fi
\def\au#1{#1} \def\ed#1{#1} \def\yr#1{#1}\def\at#1{#1}\def\jt#1{\textit{#1}}
  \def\bt#1{#1}\def\bvol#1{\textbf{#1}} \def\vol#1{#1} \def\pg#1{#1}
  \def\publ#1{#1}\def\arxiv#1{#1}\def\org#1{#1}\def\st#1{\textit{#1}}

\bibitem[Anagnostopoulos \& Bearman(1992)]{Anagnostopoulos1992ResponseNumbers}
{\sc \au{Anagnostopoulos, P.} \& \au{Bearman, P.~W.}} \yr{1992}  \at{{Response
  characteristics of a vortex-excited cylinder at low reynolds numbers}}.
  \jt{Journal of Fluids and Structures}  \bvol{6}~(1),  \pg{39--50}.

\bibitem[Batchelor(1967)]{BATCHELOR}
{\sc \au{Batchelor, G~K}} \yr{1967} {\em {An introduction to fluid
  dynamics}\/}.  \publ{Cambridge: Cambridge University Press}.

\bibitem[Bearman(2011)]{Bearman2011CircularVibrations}
{\sc \au{Bearman, P.~W.}} \yr{2011}  \at{{Circular cylinder wakes and
  vortex-induced vibrations}}.  \jt{Journal of Fluids and Structures}
  \bvol{27}~(5-6),  \pg{648--658}.

\bibitem[Bhat \& Govardhan(2013)]{Bhat2013StallNumbers}
{\sc \au{Bhat, Shantanu~S.} \& \au{Govardhan, Raghuraman~N.}} \yr{2013}
  \at{{Stall flutter of NACA 0012 airfoil at low Reynolds numbers}}.
  \jt{Journal of Fluids and Structures}  \bvol{41},  \pg{166--174}.

\bibitem[Bishop \& Hassan(1964)]{Bishop1964TheFluid}
{\sc \au{Bishop, Richard Evelyn~Donohue} \& \au{Hassan, A.Y.}} \yr{1964}
  \at{{The lift and drag forces on a circular cylinder oscillating in a flowing
  fluid}}.  \jt{Proceedings of the Royal Society of London. Series A.
  Mathematical and Physical Sciences}  \bvol{277}~(1368),  \pg{51--75}.

\bibitem[Blackburn \& Henderson(1999)]{Blackburn1999ACylinder}
{\sc \au{Blackburn, H.~M.} \& \au{Henderson, R.~D.}} \yr{1999}  \at{{A study of
  two-dimensional flow past an oscillating cylinder}}.  \jt{Journal of Fluid
  Mechanics}  \bvol{385},  \pg{255--286}.

\bibitem[Blackburn \& Karniadakis(1993)]{Blackburn1993Two-Cylinder}
{\sc \au{Blackburn, H~M} \& \au{Karniadakis, G~E}} \yr{1993}  \at{{Two- and
  Three-Dimensional Simulations of Vortex-Induced Vibration of a Circular
  Cylinder}}.  \jt{3rd Int. Offshore {\&} Polar Engng Conf., Singapore}
  ~(1977),  \pg{715--720}.

\bibitem[Brika \& Laneville(1993)]{Brika1993Vortex-inducedCylinder}
{\sc \au{Brika, D.} \& \au{Laneville, A.}} \yr{1993}  \at{{Vortex-induced
  vibrations of a long flexible circular cylinder}}.  \jt{Journal of Fluid
  Mechanics}  \bvol{250}~(EM5),  \pg{481--508}.

\bibitem[Carberry {\em et~al.\/}(2005)Carberry, Sheridan \&
  Rockwell]{Carberry2005ControlledModes}
{\sc \au{Carberry, J.}, \au{Sheridan, J.} \& \au{Rockwell, D.}} \yr{2005}
  \at{{Controlled oscillations of a cylinder: Forces and wake modes}}.
  \jt{Journal of Fluid Mechanics}  \bvol{538},  \pg{31--69}.

\bibitem[Chang(1992)]{Chang1992PotentialFlow}
{\sc \au{Chang, Chien~Cheng}} \yr{1992}  \at{{Potential flow and forces for
  incompressible viscous flow}}.  \jt{Proceedings of the Royal Society A:
  Mathematical, Physical and Engineering Sciences}  \bvol{437}~(1901),
  \pg{517--525}.

\bibitem[D'Alessio {\em et~al.\/}(1999)D'Alessio, Dennis \&
  Nguyen]{DAlessio1999UnsteadyCylinder}
{\sc \au{D'Alessio, S.~J.D.}, \au{Dennis, S.~C.R.} \& \au{Nguyen, P.}}
  \yr{1999}  \at{{Unsteady viscous flow past an impulsively started oscillating
  and translating elliptic cylinder}}.  \jt{Journal of Engineering Mathematics}
   \bvol{35}~(3),  \pg{339--357}.

\bibitem[D'Alessio \& Kocabiyik(2001)]{DAlessio2001NumericalCylinder}
{\sc \au{D'Alessio, S.~J.D.} \& \au{Kocabiyik, S.}} \yr{2001}  \at{{Numerical
  simulation of the flow induced by a transversely oscillating inclined
  elliptic cylinder}}.  \jt{Journal of Fluids and Structures}  \bvol{15}~(5),
  \pg{691--715}.

\bibitem[Davidson \& Riley(1972)]{Davidson1972JetsMotion}
{\sc \au{Davidson, B.~J.} \& \au{Riley, N.}} \yr{1972}  \at{{Jets induced by
  oscillatory motion}}.  \jt{Journal of Fluid Mechanics}  \bvol{53}~(2),
  \pg{287--303}.

\bibitem[Feng(1968)]{Feng1968TheCylinders}
{\sc \au{Feng, C.C.}} \yr{1968}  \at{{The measurement of vortex induced effects
  in flow past stationary and oscillating circular and D-section cylinders}}.
  \jt{Masters thesis, The University of British Columbia, Canada} .

\bibitem[Franzini {\em et~al.\/}(2009)Franzini, Fujarra, Meneghini, Korkischko
  \& Franciss]{Franzini2009ExperimentalCylinders}
{\sc \au{Franzini, G.~R.}, \au{Fujarra, A.~L.C.}, \au{Meneghini, J.~R.},
  \au{Korkischko, I.} \& \au{Franciss, R.}} \yr{2009}  \at{{Experimental
  investigation of Vortex-Induced Vibration on rigid, smooth and inclined
  cylinders}}.  \jt{Journal of Fluids and Structures}  \bvol{25}~(4),
  \pg{742--750}.

\bibitem[Ghias {\em et~al.\/}(2007)Ghias, Mittal \& Dong]{Ghias2007}
{\sc \au{Ghias, R.}, \au{Mittal, R.} \& \au{Dong, H.}} \yr{2007}  \at{{A sharp
  interface immersed boundary method for compressible viscous flows}}.
  \jt{Journal of Computational Physics}  \bvol{225}~(1),  \pg{528--553}.

\bibitem[Gopalakrishnan(1993)]{Gopalakrishnan1993Vortex-InducedCylinders}
{\sc \au{Gopalakrishnan, Ramnarayan}} \yr{1993}  \at{{Vortex-Induced Forces on
  Oscillating Bluff Cylinders}}. PhD thesis, Massachusetts Institute of
  Technology.

\bibitem[Govardhan \& Williamson(2000)]{Govardhan2000}
{\sc \au{Govardhan, R.} \& \au{Williamson, C.~H.K.}} \yr{2000}  \at{{Modes of
  vortex formation and frequency response of a freely vibrating cylinder}}.
  \jt{Journal of Fluid Mechanics}  \bvol{420},  \pg{85--130}.

\bibitem[Hall(1984)]{Hall1984OnFluid}
{\sc \au{Hall, Philip}} \yr{1984}  \at{{On the stability of the unsteady
  boundary layer on a cylinder oscillating transversely in a viscous fluid}}.
  \jt{Journal of Fluid Mechanics}  \bvol{146},  \pg{347--367}.

\bibitem[Hasheminejad \& Jarrahi(2015)]{Hasheminejad2015NumericalNumbers}
{\sc \au{Hasheminejad, Seyyed~M.} \& \au{Jarrahi, Miad}} \yr{2015}
  \at{{Numerical simulation of two dimensional vortex-induced vibrations of an
  elliptic cylinder at low Reynolds numbers}}.  \jt{Computers and Fluids}
  \bvol{107},  \pg{25--42}.

\bibitem[Hover {\em et~al.\/}(1998)Hover, Techet \&
  Triantafyllou]{Hover1998ForcesCrossflow}
{\sc \au{Hover, F.~S.}, \au{Techet, A.~H.} \& \au{Triantafyllou, M.~S.}}
  \yr{1998}  \at{{Forces on oscillating uniform and tapered cylinders in
  crossflow}}.  \jt{Journal of Fluid Mechanics}  \bvol{363},  \pg{97--114}.

\bibitem[Howe(1995)]{Howe1995OnNumbers}
{\sc \au{Howe, M.~S.}} \yr{1995}  \at{{On the force and moment on a body in an
  incompressible fluid, with application to rigid bodies and bubbles at high
  and low reynolds numbers}}.  \jt{Quarterly Journal of Mechanics and Applied
  Mathematics}  \bvol{48}~(3),  \pg{401--426}.

\bibitem[Kanwal(1955)]{Kanwal1955VibrationsFluid}
{\sc \au{Kanwal, R.~P.}} \yr{1955}  \at{{Vibrations of an Elliptic Cylinder and
  a Flat Plate in a Viscous Fluid}}.  \jt{ZAMM ‐ Journal of Applied
  Mathematics and Mechanics / Zeitschrift f{\"{u}}r Angewandte Mathematik und
  Mechanik}  \bvol{35}~(1-2),  \pg{17--22}.

\bibitem[Khalak \& Williamson(1999)]{Khalak1999MotionsMass-Damping}
{\sc \au{Khalak, A.} \& \au{Williamson, C.H.K.}} \yr{1999}  \at{{Motions,
  Forces and Mode Transitions in Vortex-Induced Vibrations At Low
  Mass-Damping}}.  \jt{Journal of Fluids and Structures}  \bvol{13}~(7-8),
  \pg{813--851}.

\bibitem[Kocabiyik \& D'Alessio(2004)]{Kocabiyik2004NumericalFlow}
{\sc \au{Kocabiyik, Serpil} \& \au{D'Alessio, S.J.D.}} \yr{2004}
  \at{{Numerical study of flow around an inclined elliptic cylinder oscillating
  in line with an incident uniform flow}}.  \jt{European Journal of Mechanics -
  B/Fluids}  \bvol{23}~(2),  \pg{279--302}.

\bibitem[Kumar {\em et~al.\/}(2016)Kumar, {Navrose} \&
  Mittal]{Kumar2016Lock-inCylinder}
{\sc \au{Kumar, Samvit}, \au{{Navrose}} \& \au{Mittal, Sanjay}} \yr{2016}
  \at{{Lock-in in forced vibration of a circular cylinder}}.  \jt{Physics of
  Fluids}  \bvol{28}~(11).

\bibitem[Lighthill(1986)]{Lighthill1986FundamentalsStructures}
{\sc \au{Lighthill, James}} \yr{1986}  \at{{Fundamentals concerning wave
  loading on offshore structures}}.  \jt{Journal of Fluid Mechanics}
  \bvol{173},  \pg{667--681}.

\bibitem[Magnaudet(2011)]{Magnaudet2011ANumber}
{\sc \au{Magnaudet, Jacques}} \yr{2011}  \at{{A 'reciprocal' theorem for the
  prediction of loads on a body moving in an inhomogeneous flow at arbitrary
  Reynolds number}}.  \jt{Journal of Fluid Mechanics}  \bvol{689},
  \pg{564--604}.

\bibitem[Mart{\'{i}}n-Alc{\'{a}}ntara {\em
  et~al.\/}(2015)Mart{\'{i}}n-Alc{\'{a}}ntara, Fernandez-Feria \&
  Sanmiguel-Rojas]{Martin-Alcantara2015VortexAttack}
{\sc \au{Mart{\'{i}}n-Alc{\'{a}}ntara, A.}, \au{Fernandez-Feria, R.} \&
  \au{Sanmiguel-Rojas, E.}} \yr{2015}  \at{{Vortex flow structures and
  interactions for the optimum thrust efficiency of a heaving airfoil at
  different mean angles of attack}}.  \jt{Physics of Fluids}  \bvol{27}~(7).

\bibitem[Meneghini \& Bearman(1995)]{Meneghini1995NumericalCylinder}
{\sc \au{Meneghini, J.~R.} \& \au{Bearman, P.~W.}} \yr{1995}  \at{{Numerical
  simulation of high amplitude oscillatory flow about a circular cylinder}}.
  \jt{Journal of Fluids and Structures}  \bvol{9}~(4),  \pg{435--455}.

\bibitem[Menon \& Mittal(2019)]{Menon2019}
{\sc \au{Menon, Karthik} \& \au{Mittal, Rajat}} \yr{2019}  \at{{Flow physics
  and dynamics of flow-induced pitch oscillations of an airfoil}}.  \jt{Journal
  of Fluid Mechanics}  \bvol{877},  \pg{582--613}.

\bibitem[Mittal {\em et~al.\/}(2008)Mittal, Dong, Bozkurttas, Najjar, Vargas \&
  von Loebbecke]{Mittal2008ABoundaries}
{\sc \au{Mittal, R.}, \au{Dong, H.}, \au{Bozkurttas, M.}, \au{Najjar, F.~M.},
  \au{Vargas, A.} \& \au{von Loebbecke, A.}} \yr{2008}  \at{{A versatile sharp
  interface immersed boundary method for incompressible flows with complex
  boundaries}}.  \jt{Journal of Computational Physics}  \bvol{227}~(10),
  \pg{4825--4852}.

\bibitem[Moriche {\em et~al.\/}(2018)Moriche, Flores \&
  Garc{\'{i}}a-Villalba]{Moriche2018OnNumber}
{\sc \au{Moriche, M}, \au{Flores, O} \& \au{Garc{\'{i}}a-Villalba, M}}
  \yr{2018}  \at{{On the aerodynamic forces on heaving and pitching airfoils at
  low Reynolds number}}.  \jt{J. Fluid Mech}  \bvol{828},  \pg{395--423}.

\bibitem[Morison {\em et~al.\/}(1950)Morison, Johnson \&
  Schaaf]{Morison1950ThePiles}
{\sc \au{Morison, J.R.}, \au{Johnson, J.W.} \& \au{Schaaf, S.A.}} \yr{1950}
  \at{{The Force Exerted by Surface Waves on Piles}}.  \jt{Journal of Petroleum
  Technology}  \bvol{2}~(05),  \pg{149--154}.

\bibitem[Morse \& Williamson(2009)]{Morse2009}
{\sc \au{Morse, T.~L.} \& \au{Williamson, C.~H.K.}} \yr{2009}  \at{{Prediction
  of vortex-induced vibration response by employing controlled motion}}.
  \jt{Journal of Fluid Mechanics}  \bvol{634},  \pg{5--39}.

\bibitem[Morse \& Williamson(2006)]{Morse2006EmployingVibration}
{\sc \au{Morse, T.~L.} \& \au{Williamson, C. H~K}} \yr{2006}  \at{{Employing
  controlled vibrations to predict fluid forces on a cylinder undergoing
  vortex-induced vibration}}.  \jt{Journal of Fluids and Structures}
  \bvol{22}~(6-7),  \pg{877--884}.

\bibitem[{Navrose} {\em et~al.\/}(2014){Navrose}, Yogeswaran, Sen \&
  Mittal]{Navrose2014FreeNumbers}
{\sc \au{{Navrose}}, \au{Yogeswaran, V.}, \au{Sen, Subhankar} \& \au{Mittal,
  Sanjay}} \yr{2014}  \at{{Free vibrations of an elliptic cylinder at low
  Reynolds numbers}}.  \jt{Journal of Fluids and Structures}  \bvol{51},
  \pg{55--67}.

\bibitem[Noca {\em et~al.\/}(1997)Noca, Shiels \&
  Jeon]{Noca1997MeasuringDerivatives}
{\sc \au{Noca, F.}, \au{Shiels, D.} \& \au{Jeon, D.}} \yr{1997}  \at{{Measuring
  instantaneous fluid dynamic forces on bodies, using only velocity fields and
  their derivatives}}.  \jt{Journal of Fluids and Structures}  \bvol{11}~(3),
  \pg{345--350}.

\bibitem[Noca {\em et~al.\/}(1999)Noca, Shiels \& Jeon]{Noca1999ADerivatives}
{\sc \au{Noca, F.}, \au{Shiels, D.} \& \au{Jeon, D.}} \yr{1999}  \at{{A
  comparison of methods for evaluating time-dependent fluid dynamic forces on
  bodies, using only velocity fields and their derivatives}}.  \jt{Journal of
  Fluids and Structures}  \bvol{13}~(5),  \pg{551--578}.

\bibitem[Ongoren \& Rockwell(1988)]{Ongoren1988FlowWake}
{\sc \au{Ongoren, A.} \& \au{Rockwell, D.}} \yr{1988}  \at{{Flow structure from
  an oscillating cylinder Part 2. Mode competition in the near wake}}.
  \jt{Journal of Fluid Mechanics}  \bvol{191},  \pg{225--245}.

\bibitem[Pan \& Chew(2002)]{Pan2002AFlow}
{\sc \au{Pan, L.~S.} \& \au{Chew, Y.~T.}} \yr{2002}  \at{{A general formula for
  calculating forces on a 2-D arbitrary body in incompressible flow}}.
  \jt{Journal of Fluids and Structures}  \bvol{16}~(1),  \pg{71--82}.

\bibitem[Prasanth \& Mittal(2008)]{Prasanth2008Vortex-inducedNumbers}
{\sc \au{Prasanth, T.~K.} \& \au{Mittal, S.}} \yr{2008}  \at{{Vortex-induced
  vibrations of a circular cylinder at low Reynolds numbers}}.  \jt{Journal of
  Fluid Mechanics}  \bvol{594}~(2008),  \pg{463--491}.

\bibitem[Protas {\em et~al.\/}(2000)Protas, Styczek \&
  Nowakowski]{Protas2000AnFlows}
{\sc \au{Protas, B.}, \au{Styczek, A.} \& \au{Nowakowski, A.}} \yr{2000}
  \at{{An Effective Approach to Computation of Forces in Viscous Incompressible
  Flows}}.  \jt{Journal of Computational Physics}  \bvol{159}~(2),
  \pg{231--245}.

\bibitem[Quartappelle \& Napolitano(1982)]{Quartappelle1982ForceFlows}
{\sc \au{Quartappelle, L} \& \au{Napolitano, M}} \yr{1982}  \at{{Force and
  moment in incompressible flows}}.  \jt{AIAA Journal}  \bvol{21}~(6),
  \pg{911--913}.

\bibitem[Sarpkaya(1978)]{Sarpkaya1978FluidCylinders}
{\sc \au{Sarpkaya, T.}} \yr{1978}  \at{{Fluid forces on oscillating
  cylinders}}.  \jt{Journal of the Waterway, Port, Coastal and Ocean Division}
  \bvol{104}~(3),  \pg{275--290}.

\bibitem[Sarpkaya(2001)]{Sarpkaya2001OnMorison}
{\sc \au{Sarpkaya, T.}} \yr{2001}  \at{{On the force decompositions of
  lighthill and Morison}}.  \jt{Journal of Fluids and Structures}
  \bvol{15}~(2),  \pg{227--233}.

\bibitem[Sarpkaya(2004)]{Sarpkaya2004AVibrations}
{\sc \au{Sarpkaya, T.}} \yr{2004}  \at{{A critical review of the intrinsic
  nature of vortex-induced vibrations}}.  \jt{Journal of Fluids and Structures}
   \bvol{19}~(4),  \pg{389--447}.

\bibitem[Seo \& Mittal(2011)]{Seo2011AOscillations}
{\sc \au{Seo, Jung~Hee} \& \au{Mittal, Rajat}} \yr{2011}  \at{{A
  sharp-interface immersed boundary method with improved mass conservation and
  reduced spurious pressure oscillations}}.  \jt{Journal of Computational
  Physics}  \bvol{230}~(19),  \pg{7347--7363}.

\bibitem[Singh \& Mittal(2005)]{Singh2005Vortex-inducedModes}
{\sc \au{Singh, S.~P.} \& \au{Mittal, S.}} \yr{2005}  \at{{Vortex-induced
  oscillations at low reynolds numbers: Hysteresis and vortex-shedding modes}}.
   \jt{Journal of Fluids and Structures}  \bvol{20}~(8),  \pg{1085--1104}.

\bibitem[Staubli(1983)]{Staubli1983CalculationOscillation}
{\sc \au{Staubli, T.}} \yr{1983}  \at{{Calculation of the Vibration of an
  Elastically Mounted Cylinder Using Experimental Data From Forced
  Oscillation}}.  \jt{Journal of Fluids Engineering}  \bvol{105}~(2),
  \pg{225}.

\bibitem[Triantafyllou {\em et~al.\/}(1986)Triantafyllou, Triantafyllou \&
  Chryssostomidis]{Triantafyllou1986OnCylinders}
{\sc \au{Triantafyllou, George~S.}, \au{Triantafyllou, Michael~S.} \&
  \au{Chryssostomidis, C.}} \yr{1986}  \at{{On the formation of vortex streets
  behind stationary cylinders}}.  \jt{Journal of Fluid Mechanics}  \bvol{170},
  \pg{461--477}.

\bibitem[Wang {\em et~al.\/}(2019)Wang, Zhai \&
  Chen]{Wang2019Vortex-inducedFreedom}
{\sc \au{Wang, Huakun}, \au{Zhai, Qiu} \& \au{Chen, Kaixiao}} \yr{2019}
  \at{{Vortex-induced vibrations of an elliptic cylinder with both transverse
  and rotational degrees of freedom}}.  \jt{Journal of Fluids and Structures}
  \bvol{84},  \pg{36--55}.

\bibitem[Williamson \& Govardhan(2004)]{Williamson2004Vortex-InducedVibrations}
{\sc \au{Williamson, C.H.K.} \& \au{Govardhan, R.}} \yr{2004}
  \at{{Vortex-Induced Vibrations}}.  \jt{Annual Review of Fluid Mechanics}
  \bvol{36}~(1),  \pg{413--455}.

\bibitem[Williamson \& Roshko(1988)]{Williamson1988VortexCylinder}
{\sc \au{Williamson, C. H~K} \& \au{Roshko, A.}} \yr{1988}  \at{{Vortex
  formation in the wake of an oscillating cylinder}}.  \jt{Journal of Fluids
  and Structures}  \bvol{2}~(4),  \pg{355--381}.

\bibitem[Wu(1981)]{Wu1981TheoryFlows}
{\sc \au{Wu, J.~C.}} \yr{1981}  \at{{Theory for Aerodynamic Force and Moment in
  Viscous Flows}}.  \jt{AIAA Journal}  \bvol{19}~(4),  \pg{432--441}.

\bibitem[Zdravkovich(1981)]{Zdravkovich1981ReviewShedding}
{\sc \au{Zdravkovich, M.~M.}} \yr{1981}  \at{{Review and classification of
  various aerodynamic and hydrodynamic means for suppressing vortex shedding}}.
   \jt{Journal of Wind Engineering and Industrial Aerodynamics}  \bvol{7}~(2),
  \pg{145--189}.

\bibitem[Zdravkovich(1982)]{Zdravkovich1982ModificationRange.}
{\sc \au{Zdravkovich, M.~M.}} \yr{1982}  \at{{Modification of Vortex Shedding
  in the Synchronization Range.}}  \jt{American Society of Mechanical Engineers
  (Paper)}  \bvol{104}~(December),  \pg{513--517}.

\bibitem[Zhang(2015)]{Zhang2015MechanismsInsects}
{\sc \au{Zhang, Chao}} \yr{2015}  \at{{Mechanisms for Aerodynamic Force
  Generation and Flight Stability in Insects}}. PhD thesis, Johns Hopkins
  University.

\bibitem[Zhang {\em et~al.\/}(2015)Zhang, Hedrick \&
  Mittal]{Zhang2015CentripetalInsects}
{\sc \au{Zhang, Chao}, \au{Hedrick, Tyson~L.} \& \au{Mittal, Rajat}} \yr{2015}
  \at{{Centripetal acceleration reaction: An effective and robust mechanism for
  flapping flight in insects}}.  \jt{PLoS ONE}  \bvol{10}~(8),  \pg{1--16}.

\bibitem[Zhu {\em et~al.\/}(2020)Zhu, Su \& Breuer]{Zhu2020NonlinearWing}
{\sc \au{Zhu, Yuanhang}, \au{Su, Yunxing} \& \au{Breuer, Kenneth}} \yr{2020}
  \at{{Nonlinear flow-induced instability of an elastically mounted pitching
  wing}}.  \jt{Journal of Fluid Mechanics}  \bvol{899}.

\end{thebibliography}

\end{document}